\documentclass[prx,nofootinbib,twocolumn,showkeys,groupaddress,preprintnumbers,floatfix]{revtex4-1}

\usepackage{dynlearn}
\usepackage{empheq}

\usepackage{overpic}

\newcommand{\Shannon}{\H}
\newcommand{\kB}{k_\text{B}}
\newcommand{\Wdiss}{W_\text{diss}}
\newcommand{\Wex}{W_\text{ex}}

\newcommand{\St}{\mathcal{S}}
\newcommand{\MSt}{\mathcal{M}}
\newcommand{\SSet}{\boldsymbol{\mathcal{S}}}

\newcommand{\MSet}{\boldsymbol{\mathcal{M}}}

\newcommand{\drivinghistory}[1][t]{x_{-\infty: {#1}}}
\newcommand{\drive}{x_{0:\tau}}

\newcommand{\stationary}{\boldsymbol{\pi}}
\newcommand{\GlobalEq}[1][x_0]{\boldsymbol{\pi}_{#1}}
\newcommand{\LocalEq}[2][x_0]{\GlobalEq[#1]^{(#2)}}

\newcommand{\actual}[1][0]{\boldsymbol{\mu}_{#1}}

\newcommand{\zero}{{\color{blue}{\boldsymbol{\mathtt{0}}}}}
\renewcommand{\one}{{\color{blue}{\boldsymbol{\mathtt{1}}}}}  

\newcommand{\sseq}{s_{0:\tau}}
\newcommand{\Reverse}{\protect\reflectbox{$\mathbf{R}$}}
\newcommand{\smallReverse}{\scalebox{0.7}{\Reverse}}

\newcommand{\IverL}{\bigl[ \!\! \bigl[}
\newcommand{\IverR}{\bigr] \!\! \bigr]}

\newcommand{\Left}{\text{L}}    \newcommand{\Right}{\text{R}}  

\newcommand{\NR}{\Psi}

\renewcommand{\H}{\operatorname{H}}

\usepackage{setspace}

\begin{document}

\def\ourTitle{Balancing Error and Dissipation in Computing
}

\def\ourAbstract{Modern digital electronics support remarkably reliable computing, especially
given the challenge of controlling nanoscale logical components that interact
in fluctuating environments. However, we demonstrate that the high-reliability
limit is subject to a fundamental error--energy-efficiency tradeoff that arises
from time-symmetric control: Requiring a low probability of error causes energy
consumption to diverge as logarithm of the inverse error rate for nonreciprocal
logical transitions. The \emph{reciprocity} (\emph{self}-invertibility) of a
computation is a stricter condition for thermodynamic efficiency than logical
reversibility (invertibility), the latter being the root of Landauer's work
bound on erasing information. Beyond engineered computation, the results
identify a generic error--dissipation tradeoff in steady-state transformations
of genetic information carried out by biological organisms. The lesson is that
computation under time-symmetric control cannot reach, and is often far above,
the Landauer limit. In this way, time-asymmetry becomes a design principle for
thermodynamically efficient computing. 
}

\def\ourKeywords{nonequilibrium steady state, thermodynamics, dissipation, entropy
  production, Landauer bound
}

\hypersetup{
  pdfauthor={James P. Crutchfield},
  pdftitle={\ourTitle},
  pdfsubject={\ourAbstract},
  pdfkeywords={\ourKeywords},
  pdfproducer={},
  pdfcreator={}
}

\title{\ourTitle}

\author{Paul M. Riechers}
\email{pmriechers@gmail.com}

\affiliation{Complexity Institute and School of Physical and Mathematical Sciences,
Nanyang Technological University,
637371 Singapore, Singapore}

\author{Alexander B. Boyd}
\email{abboyd@ucdavis.edu}

\affiliation{Complexity Institute and School of Physical and Mathematical Sciences,
Nanyang Technological University,
637371 Singapore, Singapore}

\author{Gregory W. Wimsatt}
\email{gwwimsatt@ucdavis.edu}

\affiliation{Complexity Sciences Center and Physics Department,
University of California at Davis, One Shields Avenue, Davis, CA 95616}

\author{James P. Crutchfield}
\email{chaos@ucdavis.edu}

\affiliation{Complexity Sciences Center and Physics Department,
University of California at Davis, One Shields Avenue, Davis, CA 95616}

\date{\today}
\bibliographystyle{unsrt}

\begin{abstract}
\ourAbstract
\end{abstract}


\preprint{\arxiv{1909.06650}}

\date{\today}
\maketitle

\renewcommand{\baselinestretch}{0.6}\normalsize
\tableofcontents
\renewcommand{\baselinestretch}{1.0}\normalsize

\setstretch{1.05}

\addtocontents{toc}{\vspace{10pt}}
\section{Introduction}

Tantalizingly, the thermodynamics of computation tells us that information
processing can be achieved with zero energy dissipation ... if one has
sufficient control over a system's microscopic degrees of freedom and can
endure the quasistatic limit of infinitely-slow processing \cite{Benn82,
Maro09, Saga14, Parr15a, Riec18}.  To be useful, though, computation must be
performed in finite time. Unfortunately, this requires additional work and
guarantees the investment is lost via dissipation. This state of affairs poses
a grand challenge to thermodynamic computing: Identify control protocols that
reliably drive a system between memory states according to a desired
computation in finite time and with minimal dissipation. Failing an answer, the
fundamental physical limits on computation remain elusive.

In point of fact, contemporary finite-time thermodynamics predicts that
energy-efficient protocols of duration $\tau$ entail dissipation that scales as
$\tau^{-1}$~\cite{Sala83,Siva12,Zulk14,Bona14,Mand16,Boyd18a}. That is,
reliable computation could be performed with arbitrarily little dissipation at
the cost of arbitrarily slow processing. 

An obvious and common implementation of finite-time computation is to ramp up a
set of forces\footnote{For example, a set of gate voltages may be applied or,
by any other method, a potential energy landscape may be altered to temporarily
couple and bias a set of initially-independent metastable potential minima.}
that transform previously-metastable memories according to the computation's
input--output map. At an appropriate later time, the forces are ramped down
(mirroring the ramp-up procedure) so that the computing system returns to its
resting state, while new memories are again stored robustly in metastable
configurations. Such implementations represent control protocols that are
symmetric in time.

Such time-symmetric protocols are common. Most notably, the primary signal
controlling information processing in contemporary microprocessors is a
time-symmetric clock-voltage signal---a several gigahertz square wave that
orchestrates all transformations of the computer's logic and memory
components~\cite{Chan92,Iyer02}.

Time-symmetric protocols are common in other settings that seek reliable
transformations. For example, recent explorations of synthetic molecular
machines externally drive molecular rotors with sinusoidal (time-symmetric)
electric fields \cite{Seld10}. And, autonomous synthetic molecular machines are
known to reliably produce steady-state rotations embedded in a
time-\emph{invariant} nonequilibrium environment~\cite{Astu17}---a special case
of time-symmetric control. They achieve this through chemical catalysis, much
as natural molecular machines operate in vivo. In vivo, reliable genetic
transformations of DNA (and reliable directional motion of molecular motors)
occur in environments with very low Reynolds number, where there are no
inertial variables to freely exert a time-asymmetric influence~\cite{Brow17,
Zhan18}. More broadly, breaking time symmetry in Brownian motion
\emph{requires} significant dissipation~\cite{Feng08}. And so, in both the
biological and engineered worlds, simple time-symmetric protocol
design---turning on and then turning off an interaction---seems to offer an
energetically-efficient and reliable way to implement change.

The goal, naturally enough, is to implement information processing in ways that
require no more than the minimal work exertion set by Landauer's
logical-irreversibility bound \cite{Land61a}: $W \geq \kB T \ln 2$. One might
even hope that this work could then be recycled rather than dissipated. Indeed,
as appreciated recently, changes in the nonequilibrium addition to free energy
can be leveraged to implement logically-irreversible computations in a
thermodynamically reversible manner~\cite{Maro09, Saga14, Parr15a, Riec18}. The
vision is of a future of hyper-efficient computing using orders-of-magnitude
less energy than currently.

However, there is a wrinkle in this optimism. The following shows that
\emph{all} time-symmetric protocols for transforming metastable memories
engender an irreducible trade-off between computational accuracy and energy
efficiency. Specifically, in the limit of highly-reliable computing with
vanishingly-small error probability $\epsilon$, the minimal dissipation under
time-symmetric controls diverges as $- \ln \epsilon$ with a coefficient
proportional to the computation's \emph{nonreciprocity}---the
nonself-invertibility of its memory-state transitions. 

That is, for reliable time-symmetric implementations of a deterministic
computation $\mathcal{C}$,  which maps each memory state $m \mapsto
\mathcal{C}(m)$, we show that the minimal work above the change in
local-equilibrium free energy is $\kB T \ln(1/\epsilon)$ whenever
$\mathcal{C}(\mathcal{C}(m)) \neq m$. That is, work $\kB T \ln(1/\epsilon)$ is
required whenever the computation iterated twice does not return to the
original memory state. Moreover, there is a correction when time-odd variables
store memory that highlights the unique advantages of both magnetic and
conformational memories.

In this way, nonreciprocity identifies a dominant cost of thermodynamic
computing for the broad class of time-symmetric control protocols. While future
work addresses more general thermodynamic implications of reliable computation,
since time-symmetric control protocols are common and often unavoidable in key
applications, they are the subject of our error-dissipation tradeoff analysis
here.

Our first step revisits the basics of dissipation in thermodynamic systems and
the cost of partial macroscopic knowledge, when the entire microstate cannot be
observed or controlled. The dissipation costs are then recast to apply to
thermodynamic computing generally and, in particular, to information storage
and processing in metastable mesoscopic states. We then analyze the role of
time symmetries in control protocols and give a thermodynamic accounting.
Dissipation scaling with error level is derived and applied in the limit of
highly-reliable computing. This is then used to analyze dissipation in erasure,
logic gates, and biological systems. We conclude, briefly comparing recent work
on control restrictions and noting directions for further exploration.
Appendices provide details underlying the theory.

\section{Computational Dissipation}

To start, consider any physical realization of a memory device with a system
Hamiltonian $\mathcal{H}_x$ parametrized by control $x \in \mathcal{X}$. For
example, $x$ could correspond to an applied electromagnetic field, a vector of
quantities specifying a potential energy landscape, or a set of piston
positions. The Hamiltonian $\mathcal{H}_x$ determines the system's
instantaneous energies $\{ E_{x}(s) \}_{s \in \SSet}$, where $\SSet$ is the set
of system microstates. We assume that the system is in contact with an
effectively memoryless heat bath at temperature $T$ that enables the system to
relax to both local and global equilibria. The instantaneous stochastic
microstate dynamics are completely determined by the system's instantaneous
Hamiltonian and the system's interaction with the heat bath.

Work is performed by driving the system via a \emph{control protocol}
$\drive=x_0 \cdots x_\tau$, which is the trajectory of the control parameters
from time $0$ to $\tau$. The work $W$ measures the accumulated change in the
energy of occupied microstates---energy that was supplied by the controller. In
the following, we are especially interested in \emph{computations} implemented
by the control protocol and in the associated work cost.

\subsection{Background}

Understanding the relationship between work costs and computations requires
tracking the energetics of microstate trajectories $s_{0:\tau}=s_0 \cdots
s_\tau$ of the system from time $0$ to $\tau$. Via the implied dynamics set 
by the time-dependent Hamiltonian $\mathcal{H}_{x_t}$ and the coupling with the
bath at temperature $T$, the initial state $s_0$ and the control protocol
$\drive$ determine the probability of microstate trajectories:
\begin{align}
\Pr\!_{\drive} (\St_{0:\tau}& =\sseq|\St_0=s_0)  \nonumber \\
   & \equiv \Pr(\St_{0:\tau}  =\sseq|\St_0=s_0, X_{0:\tau}=x_{0:\tau}) 
  ~,
\label{fig:trajectory}
\end{align}
where $\St_t$ is the random variable for the microstate at time $t$ and
$\St_{0:\tau}$ is the random variable chain for the full state trajectory. The
subscripted probability $\Pr_{\drive} (\cdot)$ denotes the probability
distribution \emph{induced} by the driving protocol $\drive$. This is the same
as the probability \emph{conditioned} on the driving protocol, as described by
Eq.~\eqref{fig:trajectory}.

Hamiltonian control implies microscopic reversibility of the instantaneous
system--bath dynamics. This means state trajectories that release energy into the
heat bath ($Q > 0$) are exponentially more likely than the time-reversed
microstate trajectory of the system (that absorb heat), if the protocol were run in
reverse~\cite{Croo99,Jarz00}. In the Markovian limit:\footnote{In the more
general non-Markovian case, Eq.~\eqref{eq:CrooksLemma} becomes an
\emph{in}equality which nevertheless still upholds our main results bounding
computational dissipation.}
\begin{align}
\frac{\Pr_{\drive} \bigl( \St_{0:\tau} = \sseq | \St_0 = s_0 \bigr) }{\Pr_{ \smallReverse(\drive)} \bigl( \St_{0:\tau} = \Reverse(\sseq) | \St_0 = s_\tau^\dagger \bigr)} = e^{\beta Q{(\sseq, \drive)}}
  ~,
\label{eq:CrooksLemma}
\end{align}
where $\beta = ( \kB T )^{-1}$ and $\kB$ is Boltzmann's constant. $\Reverse$ 
denotes \emph{time-reversal}, including not only reversal of the time ordering,
but also conjugation of time-odd variables such as momentum and spin: $s_0 \dots
s_\tau \mapsto s_\tau^\dagger \dots s_0^\dagger$. For example, if the
microstate $s = (\vec{q}, \vec{\wp} )$ is a collection of spatial $\vec{q}$ and
momentum $\vec{\wp}$ degrees of freedom, then the conjugation simply flips all
of the momentum degrees of freedom $s^\dagger = (\vec{q}, - \vec{\wp})$.

Thermodynamic irreversibility is quantified by \emph{entropy production}
$\Sigma$ which, by the second law, is nonnegative on average: $\braket{\Sigma} \geq 0$. In
our setting, $\Sigma$ is simply the net change in the component entropies of
both the system and the bath.

More specifically, entropy production decomposes into the change in heat-bath
entropy $Q(s_{0:\tau},x_{0:\tau}) / T$ beyond any compensating reduction in the
system's microstate entropy. The system's instantaneous microstate entropy is
given by the nonequilibrium \emph{surprisal} $- \kB \ln \actual[t](s_t) $,
where $\actual[t](s_t) = \Pr_{\drivinghistory}(\St_t=s_t)$ is the current
microstate's probability given the \emph{entire history} of preparation and
driving \cite{Croo99, Seif05, Espo11a}.\footnote{Hence, $\actual[t](s_t)$ is
the expected probability of being in the actual microstate $s_t$ over many
trials where the system is prepared and driven in exactly the same way. The
fact that this distribution is not $\delta$-distributed is due to the
stochasticity induced by the interaction with the heat bath (and any other
uncontrollable degrees of freedom).} Accordingly, the entropy production is:
\begin{align*}
\Sigma =
  \frac{1}{T} Q{(\sseq, \drive)} +  \Delta \bigl( - \kB \ln \actual[t](s_t) \bigr)
  ~,
\end{align*}
where $\Delta$ indicates the change from time $0$ to $\tau$. Specifically,
$\actual[0]$ is the distribution over microstates given the system's initial
preparation and $\actual[\tau]$ is the time-evolved version of $\actual[0]$
under the influence of the driving $\drive$.

This results in the trial-specific entropy production~\cite{Croo99,Jarz00}: 
\begin{align}
&\Sigma  = \frac{1}{T} Q(s_{0:\tau},x_{0:\tau})+ \kB \ln \frac{\actual[0](s_0)}{\actual[\tau](s_\tau)} \nonumber
\\ & = \kB \ln \left(  
\frac{\Pr_{\drive} \bigl( \St_{0:\tau} = \sseq | \St_0 = s_0\bigr) \actual[0](s_0)}{\Pr_{\smallReverse(\drive)} \bigl( \St_{0:\tau} = \Reverse(\sseq) | \St_0 = s_\tau^\dagger \bigr)\actual[\tau](s_\tau)}
\right) \nonumber
\\ & = \kB  \ln \left(  
\frac{\Pr_{\drive} \bigl( \St_{0:\tau} = \sseq | \St_0 \sim \actual[0] \bigr) }{\Pr_{\smallReverse(\drive)} \bigl( \St_{0:\tau} = \Reverse(\sseq) | \St_0 \sim \actual[\tau]^\dagger \bigr)}
\right)
  ~,
\label{eq:WdissGeneralMicro}
\end{align}
where $\actual[\tau]^\dagger$ is the distribution such that
$\actual[\tau]^\dagger(s) = \actual[\tau](s^\dagger)$. The condition $\St_0
\sim \boldsymbol{\mu}$ means the random variable $\St_0$ is distributed
according to $\boldsymbol{\mu}$, as in Ref.~\cite{Cove06a}. We used the fact
that $\actual[\tau](s) = \actual[\tau]^\dagger(s^\dagger)$ in going from the
second to the third line.

When all influences driving the system away from equilibrium are controlled
changes to the system Hamiltonian, the total entropy production is proportional
to the \emph{dissipated work} $\Wdiss$. This is the accumulation of lost
opportunities to extract work:
\begin{align*}
W_\text{diss} & = T \Sigma
  ~.
\end{align*}
The entropy production and dissipated work both measure a given computation's
efficiency, as they both quantify the degree to which more work was done than
necessary by an unrestricted controller.

The \emph{expected} dissipated work then quantifies the difference between the
average amount of work done beyond the change in nonequilibrium free
energy~\cite{Espo12}:
\begin{align}
\langle W_\text{diss} \rangle = \langle W \rangle -\Delta \mathcal{F} ~.
\label{eq:WvsWdiss}
\end{align} 
This is the average work that has been irreversibly lost, since the
nonequilibrium free energy $\mathcal{F}$ is the expected amount of energy that
possibly could be extracted as work ~\cite{Parr15a}.

Calculating the dissipated work from these two quantities seems to require
explicit knowledge of the Hamiltonian, since the work is the integrated change
in the energy due to changes in control: 
\begin{align*}
\langle W \rangle = \sum_s \int_{0}^{\tau} dt \, 
    \actual[t](s) \,
    \frac{\partial E_{x_t}(s)}{\partial x_t} 
    \frac{dx_t}{dt}
	~,
\end{align*}
and the nonequilibrium free energy is the difference between the average energy
and the microstate entropy \cite{Parr15a}:
\begin{align*}
\mathcal{F}(t)
  = \langle \mathcal{H}_{x_t} \rangle - \kB T \Shannon ( \mathcal{S}_t )
  ~,
\end{align*}
where $\Shannon (Z) = - \sum_{z} \Pr(Z=z) \ln \Pr(Z=z)$ is the Shannon entropy
of the driven system in nats.\footnote{In point of fact, knowledge of the
Hamiltonian is not strictly necessary if one instead calculates $\braket{
\Wdiss } = \braket{ \Wex } - \Delta \mathcal{F}_\text{add}$, where $\Wex = W -
\Delta F_{x}^{\text{eq}}$ and $\mathcal{F}_\text{add} = \kB T \DKL{ \actual[t]
}{ \boldsymbol{\pi}_{x_t} }$. With knowledge (or observation) of the
equilibrium distribution $\boldsymbol{\pi}_{x}$ associated with each control
setting $x$, one can leverage the Boltzmann relationship
$\boldsymbol{\pi}_{x}(s) = e^{- \beta [ E_{x}(s) - F_{x}^{\text{eq}} ]}$ to
calculate: $\beta \braket{ \Wex } = \sum_s \int_0^\tau dt \, \actual[t](s) \,
\frac{\partial}{ \partial x_t } \bigl( - \ln  \boldsymbol{\pi}_{x_t}(s) \bigr)
\frac{d x_t}{ dt }$. Still, Eq.~\eqref{eq:AverageDissipation} presents an even
more tempting opportunity for calculating $\braket{ \Wdiss }$ directly from
observed trajectory probabilities.}  

However, due to Eq.~\eqref{eq:WdissGeneralMicro}, we can calculate the
dissipated work for a control protocol using only the probability of forward
trajectories under forward driving $\rho(\St_{0:\tau}=s_{0:\tau})=\Pr_{\drive}
\bigl( \St_{0:\tau} = \sseq | \St_0 \sim \actual[0] \bigr)$ and reverse
microstate trajectories under reverse driving
$\rho^R(\St_{0:\tau}=s_{0:\tau})=\Pr_{\smallReverse(\drive)} \bigl(
\St_{0:\tau} = \Reverse(\sseq) | \St_0 \sim \actual[\tau]^\dagger \bigr)$.
Averaging over forward trajectories produces a relative entropy:
\begin{align}
\beta \langle W_\text{diss} \rangle &= \sum_{s_{0:\tau}} \rho(\St_{0:\tau}=s_{0:\tau}) \ln \frac{\rho(\St_{0:\tau}=s_{0:\tau})}{\rho^R(\St_{0:\tau}=s_{0:\tau})} \nonumber
\\ & = \DKL{ \rho(\St_{0:\tau}) }{ \rho^R(\St_{0:\tau}) } ~,
\label{eq:AverageDissipation}
\end{align}
where $\DKL{\cdot}{\cdot}$ is the Kullback--Leibler divergence.

Eq.~\eqref{eq:AverageDissipation} appears superficially similar to many previous related results,
including Refs.~\cite{Jarz06a,Kawa07,Gome08a,Rold10b,Horo09a}.  
However these previous results all required that the system starts in a
steady-state distribution (either equilibrium or a nonequilibrium steady state),
which precludes the storage of general initial memories.
Eq.~\eqref{eq:AverageDissipation} overcomes that previous limitation
via its dependence on $\actual$ and $\actual[\tau]^\dagger$---a crucial generalization that will allow us to 
address the thermodynamics of general memory transformations.

By expressing the dissipated work and entropy
production in terms of the relative entropy between forward and reverse
trajectories, Eq.~\eqref{eq:AverageDissipation} provides a method to reconstruct a system's thermodynamics
without explicit knowledge of the underlying Hamiltonian mechanics. This
only requires knowledge of trajectory probabilities under forward and reverse
driving and these can be determined experimentally.
However, practical considerations can stand in the way: Often direct
observations of a thermodynamic system's underlying microstate trajectories are
not available. 

Fortunately, via the information processing inequality~\cite{Cove06a},
this powerful relation leads to bounds on dissipated work based on
\emph{observed trajectories}, which contain only partial details of the full
microstate trajectory. 
In particular, any observable coarse-graining of the microstate trajectory
yields a lower bound on the dissipation~\cite{Kawa07,Gome08a,Rold10a}.
This will allow us to infer new lower bounds on computational dissipation
from only partial knowledge related to the logical dynamics of the computation itself.

\subsection{Thermodynamic Computing}

\begin{figure}
\centering
\includegraphics[width=.8\columnwidth]{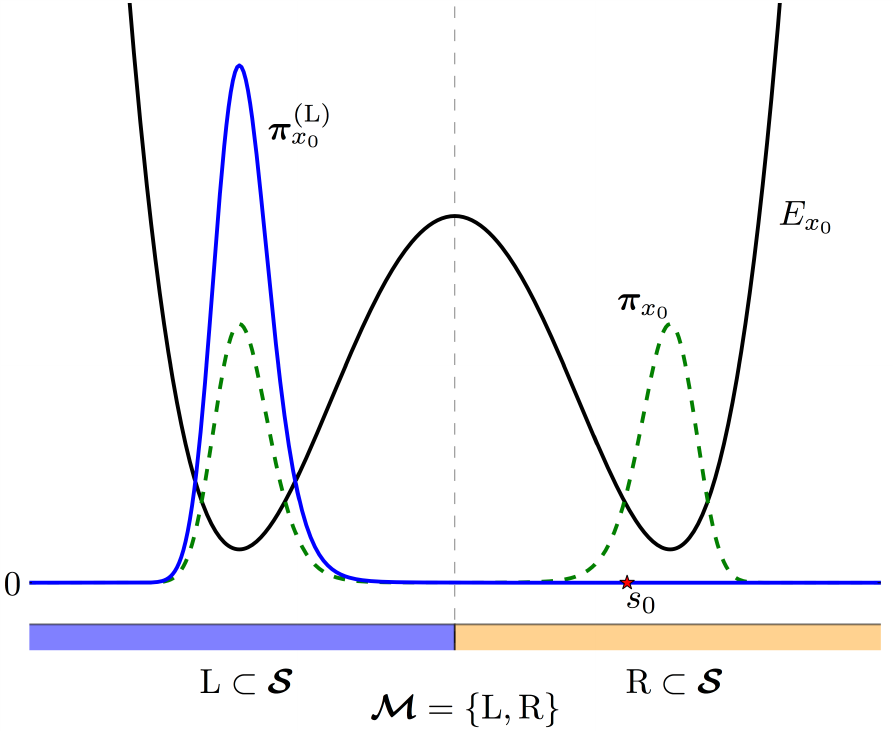}
\caption{Bistable memory element: A cross-section (at constant momentum $\wp_0$) of
	both global and local equilibrium. Initial system state $s_0 = (q_0, \wp_0)$ (red star) lies in a continuum. 
	The energy landscape (black curve) has two symmetric minima that effectively partition the space into robust
	Left (blue) and Right (yellow) spatial memory states. The stable equilibrium
	distribution (green dashed curve) is uniformly distributed over the Left
	and Right states. The restriction of the equilibrium distribution to the
	Left well (blue curve) is metastable, only locally in equilibrium.
  }
\label{fig:MetastableMemory} 
\end{figure}

To be usefully manipulated, memories must be physically addressable.
Accordingly, we define the set $\MSet$ of \emph{memory states} as a partition
of the set $\SSet$ of microstates, such that each memory state $m \in \MSet$ is
a subset of the microstates: $m \subset \SSet$ and $\bigcup m = \SSet$. In this
way, the memory state random variable $\MSt_t$ at time $t$ is determined by the
microstate $\St_t$. Moreover, for memory states to robustly hold memories
between computations, we assume that partitioning is such that each memory
state corresponds to a state-space region that is effectively autonomous when
no computation is being performed, resulting in metastability of the
memory.\footnote{\emph{During} a computation, though, driving couples memory
states to instantiate nontrivial computing.} Figure \ref{fig:MetastableMemory}
shows a spatial system that has been partitioned into Left ($\Left$) and Right
($\Right$) memory states, each of which corresponds to a minimum in the energy
landscape.

\begin{figure}
\centering
\includegraphics[width=0.8\columnwidth]{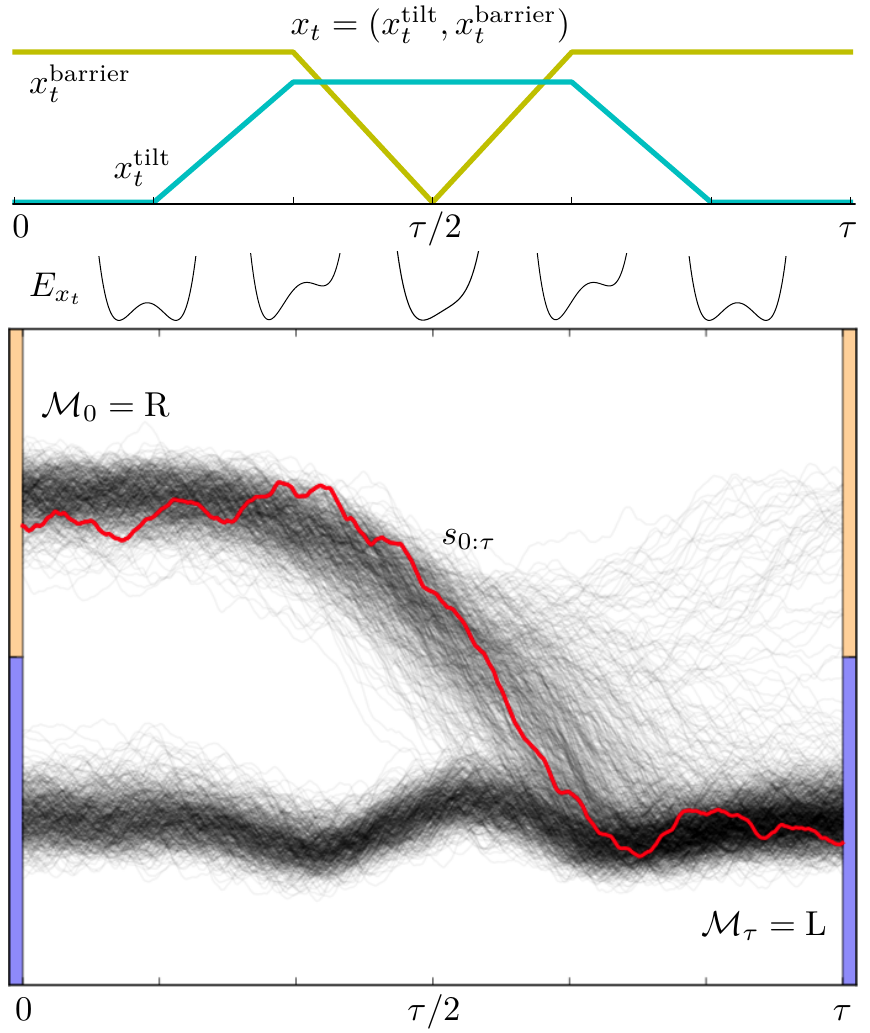}
\caption{(Top) Time-symmetric control implementing reliable information
	processing: erasure to $\Left$. (Middle) Controlled potential
	$E_{x_t}$ at several times during the erasure protocol. (Bottom)
	Representative behavior via $1000$ microstate trajectories in gray which
	sketches a profile of $\actual[t]$. A particular microstate trajectory
	$s_{0:\tau}$ is highlighted in red. (Note that $q_{0:\tau}$, shown
	explicitly, implies $\wp_{0:\tau}$ by differentiation and so also implies
	$s_{0:\tau}$.)
	}
\label{fig:MetastableComputing} 
\end{figure}

Partitioning microstates into metastable memory states $\MSet$ introduces an
observational channel for monitoring computation. The task of thermodynamic
computation is to map an initial memory state $m$ to a final memory state $m'$
according to a given conditional probability distribution:
\begin{align*}
p(m \rightarrow m') = \Pr(\mathcal{M}_\tau = m'| \mathcal{M}_0=m)
  ~.
\end{align*}
For instance, as shown in Fig.~\ref{fig:MetastableComputing} for
Fig.~\ref{fig:MetastableMemory}'s system, erasure is composed of two desired
transitions with high probabilities---$p(\Left \rightarrow \Left) \approx 1$
and $p(\Right \rightarrow \Left) \approx 1$---between the two metastable memory
states. (The latter are labeled $\Left$ and $\Right$ according to the common
use of left and right minima in a double-well potential landscape to store a
bit \cite{Jun14a}.) A computation occurs due to a particular stochastic
trajectory, such as the red path shown in Fig. \ref{fig:MetastableComputing}.
Naturally, the statistics of the memory state transition probability $p(\Left
\rightarrow \Left)$, say, come from the ensemble of trajectories. While
stochasticity is the rule in thermodynamic computations, the following focuses
on nearly-deterministic computations, such as highly-reliable erasure.

To lower-bound the thermodynamic dissipation associated with computation, we consider the coarse graining $s_{0:\tau} \mapsto (m, m')$, where $s_0 \in m$ and $s_\tau \in m'$. This maps the detailed microstate trajectory to the logical transformation it induces between initial and final memory states. For a particular realization $\MSt_0 = m$ and $\MSt_\tau = m'$, the time-reversal of the initial and final memory states is: $\Reverse(m, m') = (m'^\dagger, m^\dagger)$, where the conjugate $m^\dagger \equiv \{ s^\dagger : s \in m\}$.  This leads to observable memory transitions: 
\begin{align}
\rho \bigl( (\MSt_{0}, & \MSt_\tau) = (m, m') \bigr) \nonumber \\
  & = \sum_{s_{0:\tau}} \delta_{s_0 \in m} \delta_{s_\tau \in m'} \,
  \rho( \St_{0:\tau} = \sseq) \nonumber \\
  & = 
  \Pr\!_{\drive} \bigl( \MSt_0 = m, \MSt_\tau = m' | \St_0 \sim \actual[0] \bigr) 
\end{align}
and the reversal probabilities:
\begin{align}
\rho^R \bigl( & (\MSt_{0}, \MSt_\tau) = (m, m') \bigr) \nonumber \\
  & = \sum_{s_{0:\tau}} \delta_{s_0 \in m} \delta_{s_\tau \in m'} \,
  \rho^R( \St_{0:\tau} = \sseq) \nonumber \\
  & = \Pr\!_{\smallReverse (\drive)} \bigl( \MSt_0= m'^\dagger , \MSt_\tau= m^\dagger | \St_0 \sim \actual[\tau]^\dagger \bigr)
  ~.
\end{align}

Appendix \ref{sec:CoarseGrainingDerivation_Computation} uses this trajectory
coarse-graining to show that the dissipated work is lower-bounded by a function
of the net transition probabilities between memory states:
\begin{align}
\beta \braket{\Wdiss} 
  & = \DKL{ \rho(\St_{0:\tau}) }{ \rho^R(\St_{0:\tau}) }
  \label{eq:avgWdissGeneralMicro} \\
  & \geq \DKL{ \rho(\MSt_{0}, \MSt_\tau) }{ \rho^R(\MSt_{0}, \MSt_\tau)  }
  \label{eq:avgWdissGeneralMacro} \\
  & = \Delta \Shannon(\MSt_t) 
  + \sum_{m, m' \in \MSet} \actual[0](m) \, d(m,m')
  \label{eq:WdissDecomp}
  ~,
\end{align}
where $\actual[t]$ is the same preparation-and-driving-induced probability
measure as previously introduced, such that $\actual[t](m) =  \Pr(\MSt_t = m  |
\St_t \sim \actual[t]) = \sum_{s \in m} \actual[t](s)$, and:
\begin{align*}
\Delta \Shannon(\MSt_t) =  \sum_{m \in \MSet} \Bigl( \actual[0] (m) \ln
\actual[0] (m) - \actual[\tau] (m) \ln \actual[\tau] (m) \Bigr)
\end{align*}
is the change in Shannon entropy (in nats) of the coarse-grained memory states. 

Note that Eq.~\eqref{eq:WdissDecomp} defined:
\begin{align}
d(m,& m') \equiv 
\Pr_{\drive} \bigl(  \MSt_\tau = m' \big| \St_0 \sim \actual[0]^{(m)} \bigr) 
  \nonumber \\
  & \times
  \ln \left( \frac{\Pr_{\drive} \bigl(  \MSt_\tau = m'  \big| \St_0 \sim \actual[0]^{(m)}  \bigr) }{\Pr_{ \smallReverse (\drive)} \bigl( \MSt_\tau= m^\dagger \big| \St_0 \sim \actual[\tau]^{\dagger (m'^\dagger)}  \bigr) } \right)
  ,
\label{eq:dmmDef}
\end{align}
where $\actual[t]^{(m)} \equiv \delta_{\St_t \in m} \actual[t] / \actual[t](m)$
is the renormalized microstate distribution $\actual[t]$ restricted to memory
$m$'s microstates. This compares (i) the probability of transitioning from
memory state $m$ to $m'$ to (ii) the probability of returning to $m^\dagger$
upon subsequent momentum-conjugation of the microstate distribution in $m'$ and
reversal of the control protocol. One concludes that dissipation is due to
statistical irreversibility. At first sight, Eqs.~\eqref{eq:WdissDecomp} and
\eqref{eq:dmmDef} do not appear to significantly simplify the problem of
inferring dissipation from partial knowledge. 
However, the practical constraints of reliable
computation greatly simplify $d(m,m')$, as we will see shortly.

\begin{figure}
\centering
\includegraphics[width=0.78\columnwidth]{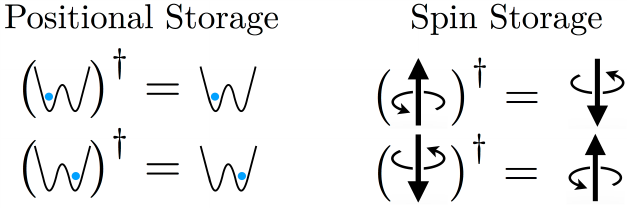}
\caption{Common memory elements are either time-reversal invariant (where
	$m^\dagger = m$) or flip under time-reversal (such that $m^\dagger \neq m$
	although $m^\dagger \in \MSet$).
	}
\label{fig:ConjugateMemory} 
\end{figure}

To emphasize, written as either Eq.~\eqref{eq:avgWdissGeneralMacro} or
\eqref{eq:WdissDecomp}, the bound applies when the conjugate $m^\dagger \equiv
\{ s^\dagger : s \in m\}$ of a memory state $m$ is an element of the memory
partition, $m^\dagger \in \MSet$, for all $m$. (However, see
App.~\ref{sec:CoarseGrainingDerivation_Computation} for the more general case.)
This occurs, for example, when spatial cells store a memory that is indifferent
to rapidly fluctuating momentum degrees of freedom, so that $m = m^\dagger$. A
similar simplification occurs for magnetic memory where $s$ and $s^\dagger$ are
always in different memory states, such that $m^\dagger \in \MSet$ although $m
\neq m^\dagger$. These two types of memory, illustrated in Fig.~\ref{fig:ConjugateMemory}, 
lead to notable physical consequences; ones that we
explore in a sequel.

Equation~\eqref{eq:WdissDecomp} is close to yielding a bound on entropy
production divorced from knowledge of microstate dynamics. $\Pr_{\drive}(
\MSt_\tau | \St_0 \sim \actual[0]^{(m)})$, for instance, can be recovered from
observations, given a system's initialization $\actual[0]^{(m)}$ and the
driving $\drive$. However, for $\Pr_{\smallReverse(\drive)} \bigl( \MSt_\tau |
\St_0 \sim \actual[\tau]^{\dagger (m)} \bigr)$, it is unclear how to
experimentally prepare the distribution $\actual[\tau]^{\dagger (m)}$ in the
presence of time-odd variables, like momentum. Fortunately, as we now show,
this obstacle can be removed for thermodynamic computations using metastable
memories. Ignoring the underlying microstate dynamics and, instead, only
tracking the logical dynamics, we bound the work production while computing.

\subsection{Metastable Processing} 

Further simplifying Eq. \eqref{eq:WdissDecomp}'s bound on dissipated work turns
on recognizing a common property of information processing systems: They hold
memory states as metastable nonequilibrium distributions. This has profound
thermodynamic implications.

The implied timescale separation requires that relaxing to the
local-equilibrium distribution $\LocalEq[x]{m}$ within each memory region is
fast, while relaxing to the global equilibrium distribution $\GlobalEq[x]$ is
much slower than the timescale of (and between) a system's computational steps.
On the one hand, \emph{metastable distributions} are those that correspond (up
to some approximation) to a normalized weighted sum of local-equilibrium
distributions. On the other, the \emph{global equilibrium distribution}
corresponding to any $x$ is the canonical one: $\GlobalEq[x](s) = e^{- \beta
E_x(s)} / \sum_{s' \in \SSet} e^{- \beta E_x(s')}$. The system approaches it in
the limit of infinite time, if the control setting $x$ is held fixed. In this,
the \emph{local-equilibrium distributions} $\LocalEq[x]{m}(s) = \delta_{s \in
m} \GlobalEq[x](s)/ \sum_{s' \in m} \GlobalEq[x](s')$ are the canonical
distributions, associated with each nearly-autonomous region of state space,
that local densities approach much more quickly due to timescale
separation~\cite{Espo12}.

If metastable memories are stored robustly between computations, then the
memory-state distribution $\actual[t](m)$ at times $t=0$ and $t=\tau$ contain
almost all the information about the microstate distributions at the start
and end times; or, at least, at the beginning and shortly \emph{after} the end
time. Indeed, the local distributions associated with each memory state are
then nearly local-equilibrium distributions. If, to the contrary, the ending
microstate distribution $\actual[\tau](s)$ is \emph{not} yet metastable, then
it quickly \emph{relaxes} to the metastable distribution $\actual[\tau + \delta
t] (s) \approx \sum_{m \in \MSet} \actual[\tau](m) \, \LocalEq[x_\tau]{m} \!
(s)$. This results in yet more dissipation.

To include all dissipation associated with a computation, we extend a given
protocol's start and end times to include most of the post-computation
relaxation to metastability. Assuming a metastable distribution at the
beginning and end of the computation means $\actual^{(m)} \approx
\LocalEq[x_0]{m}$ and $\actual[\tau]^{(m)} \approx \LocalEq[x_\tau]{m}$
and the full microstate distributions are then $\actual[0] \approx \sum_{m \in
\MSet} \actual[0](m) \LocalEq[x_0]{m}$ and $\actual[\tau] \approx \sum_{m \in
\MSet} \actual[\tau](m) \LocalEq[x_\tau]{m}$, respectively. Thus, the beginning
and ending microstate distributions are almost entirely determined by the
distribution over memory states.

Initial and final memory-system metastability means that the
empirically-observable memory transitions are always implicitly conditioned on
metastability:
\begin{align*}
\Pr_{\drive} & \bigl(  \MSt_\tau = m'  \big| \MSt_0 = m \bigr)  \nonumber \\
  &= 
    \Pr_{\drive} \bigl(  \MSt_\tau = m'  \big| \MSt_0 = m , \St_0 \sim \LocalEq[x_0]{m}  \bigr)
	~.
\end{align*}
Similarly, if we operate the control protocol in reverse, starting metastably
under the influence of $x_\tau^\dagger$, the observed memory-transition
probabilities are:
\begin{align}
& \Pr_{\smallReverse(\drive)} \bigl(  \MSt_\tau = m^\dagger  \big| \MSt_0 = m'^\dagger \bigr) \nonumber \\
   & \quad =
         \Pr_{\smallReverse(\drive)} \bigl(  \MSt_\tau = m^\dagger  \big| \MSt_0 = m'^\dagger , \St_0 \sim \boldsymbol{\pi}_{x_\tau^\dagger}^{(m'^\dagger)}   \bigr) ~.
\label{eq:ImplicitReverseMemoryConditioning}         
\end{align}
Note, for comparison with Eq.~\eqref{eq:dmmDef}, that:\footnote{That $(\boldsymbol{\pi}_{x})^\dagger = \boldsymbol{\pi}_{x^\dagger}$ follows from the energy eigenvalues being invariant under time reversal: $E_x(s) = E_{x^\dagger}(s^\dagger)$. Note that equilibrium probability depends only on the energy of the state and on the partition function. $E_x(s) = E_{x^\dagger}(s^\dagger)$ implies $Z_x = Z_{x^\dagger}$. These properties together therefore imply $(\boldsymbol{\pi}_{x})^\dagger = \boldsymbol{\pi}_{x^\dagger}$.}
\begin{align*}
\boldsymbol{\pi}_{x}^\dagger (s) = \boldsymbol{\pi}_{x}(s^\dagger) =
\boldsymbol{\pi}_{x^\dagger}(s)
  ~.
\end{align*}
And so, we have:
\begin{align*}
\boldsymbol{\pi}_{x_\tau^\dagger}^{(m'^\dagger)}  =
\boldsymbol{\pi}_{x_\tau}^{\dagger (m'^\dagger)}  = \actual[\tau]^{\dagger
(m'^\dagger)}
  ~.
\end{align*}

This allows us to remove all dependence on microstates from the work bound
and estimate dissipation purely from observed memory-state trajectories.  
In particular, Eq.~\eqref{eq:dmmDef} simplifies to: 
\begin{align}
d(m,m')&=\Pr_{\drive}(\MSt_\tau=m'|\MSt_0=m) \nonumber
\\ & \times \ln \frac{\Pr_{\drive}(\MSt_\tau=m'|\MSt_0=m)}{\Pr_{\smallReverse(\drive)}(\MSt_\tau=m^\dagger|\MSt_0=m'^\dagger)}
  ~.
\label{eq:dmm_meqmdag}
\end{align}
Now, the dissipated work bound can be expressed in terms only of the
probabilities of memory-state inputs and outputs:
\begin{align*}
\beta \langle W_\text{diss} \rangle & \geq  
\Delta \Shannon(\MSt_t) + 
\sum_{m,m'} \Pr_{\drive}(\MSt_\tau=m' , \MSt_0=m) \\
  & \qquad \times \ln \frac{\Pr_{\drive}(\MSt_\tau=m' | \MSt_0=m)}{\Pr_{\smallReverse(\drive)}(\MSt_\tau=m^\dagger | \MSt_0=m'^\dagger)}
  ~.
\end{align*}
This bounds a computation's dissipated energy using only knowledge of the
memory-state transitions that result from forward driving $\drive$ and reverse
driving $\Reverse(\drive)$. However, additional simplifications arise, if the
driving is time-symmetric.

\subsection{Time Symmetric Driving}

Let us now consider the consequences for reliable computing with
\emph{time-symmetric} protocols---those for which $\drive =
\Reverse(\drive)$---as shown in the upper panel of Fig.~\ref{fig:MetastableComputing}'s
computation. In these cases, Eq.~\eqref{eq:dmm_meqmdag} simplifies
considerably:
\begin{align}
& d(m,m') =
\Pr_{\drive} \bigl(  \MSt_\tau = m'  \big| \MSt_0=m \bigr) \nonumber \\
  & \qquad \times \ln \left( \frac{\Pr_{\drive} \bigl(  \MSt_\tau = m'  \big| \MSt_0=m \bigr) }{\Pr_{\drive} \bigl( \MSt_\tau = m^\dagger \big| \MSt_0=m'^\dagger  \bigr) } \right) ~.
\label{eq:dmmSimplified}
\end{align}
With Eqs.~\eqref{eq:WdissDecomp} and \eqref{eq:dmmSimplified}, we arrive at the
remarkable result that the minimal dissipation depends only on 
memory-transition probabilities actually exercised by the computation: 
\begin{align*}
d(m,m') = p(m \to m') \ln \frac{p(m \to m')}{p(m'^\dagger \to m^\dagger)}
  ~,
\end{align*}
where we rewrote the result suggestively denoting the computation as $p(m \rightarrow m')=\Pr_{\drive}(\MSt_\tau=m'|\MSt_0=m)$.
(Conveniently, no longer is there a dependence on counterfactual 
probabilities that \emph{would} be induced by time-reversed driving.)
Thus, the bound on dissipated work $\langle W_\text{diss}
\rangle^{t-\text{sym}}_\text{min}$ includes a term $\Delta H(\MSt)$ that
depends on initial and final observable distributions and one that depends on
the transition paths between memory states: 
\begin{empheq}{align}
&\beta \langle W_\text{diss} \rangle^{t\text{-sym}}_\text{min}
  = \Delta H(\MSt_t) \nonumber \\
  & \qquad + \sum_{m,m'}\actual[0](m)p(m \rightarrow m')
  \ln \frac{p(m \rightarrow m')}{p(m'^\dagger \rightarrow m^\dagger)}
  ~.
\label{eq:time symmetric}
\end{empheq}
This demonstrates how restricting to time-symmetric control leads to
unavoidable dissipation that depends explicitly on a computation's logic. 

This adds to a short list of work bounds determined by the computation
implemented. Original on this list---the minimal work required for a
computation without control restrictions---is simply Landauer's bound:
\begin{align*}
\langle W \rangle^\text{Landauer}_\text{min} \equiv - \kB T \Delta \Shannon(\MSt_t)
  ~,
\end{align*}
when memory states all have equal local free energies. More generally, though,
as seen from Eq.~\eqref{eq:WvsWdiss}, the minimal work required for a
computation is: 
\begin{align}
\braket{W}_\text{min} & \equiv \Delta \mathcal{F} \\
  & = \Delta  \braket{ F_{x_t}^{(\MSt_t)}}
  + \langle W \rangle^\text{Landauer}_\text{min}
  ~,
\end{align} 
which is achieved in the limit of zero dissipation.
Here, we recalled that for metastable distributions the nonequilibrium free
energy decomposes as:
\begin{align*}
\mathcal{F} = \braket{ F_{x_t}^{(\MSt_t)}} - \kB T \Shannon(\MSt_t)
  ~,
\end{align*}
where $F_{x}^{(m)} \equiv - \kB T \ln \bigl( \sum_{s \in m}
e^{-\beta E_x(s)} \bigr)$ is the local-equilibrium free energy associated with
the memory $m$~\cite{Parr15a, Riec18}. 

An important related but often overlooked result emerges. Even for
logically-irreversible computations, the minimal work required by this
generalized Landauer bound is \emph{reversible}: it corresponds to a stored
change in nonequilibrium free energy that can be retrieved. Thus, ultimately
no work needs to be dissipated. We \emph{could} have $\braket{\Wdiss} = 0$ in
the case of unrestricted control. And, this alerts us to additional
opportunities for optimizing implementations.

Appealing to Eq.~\eqref{eq:WvsWdiss} again, we see that the minimum total work
under time-symmetric driving must be:\footnote{Common computational substrates
have $\Delta \braket{ F_{x_t}^{(m)}} = 0$, since engineered memories typically
have equal local-equilibrium energies. The change in local free energy is
nevertheless likely important in the manipulation of biological memories.}
\begin{widetext}
\begin{align}
\langle W \rangle^{t\text{-sym}}_\text{min}
  & =  \Delta  \braket{ F_{x_t}^{(\MSt_t)}}
  + \langle W \rangle^\text{Landauer}_\text{min}
  + \langle W_\text{diss} \rangle^{t\text{-sym}}_\text{min} 
  \nonumber \\
  & ~ =  \Delta \braket{ F_{x_t}^{(\MSt_t)}}
  + \kB T \sum_{m,m'}\actual[0](m)p(m \rightarrow m')
  \ln \frac{p(m \rightarrow m')}{p(m'^\dagger \rightarrow m^\dagger)}
  ~.
  \label{eq:WorkRequired}
\end{align}
\end{widetext}

The consequences are immediate. Computing with time-symmetric protocols
requires additional work above and beyond the Landauer bound. Moreover, all of
this work---beside the reversible Landauer contribution and the change in local
free energy---must be irreversibly dissipated since it is not stored as nonequilibrium free energy.
$\braket{\Wdiss} > 0$, quantified precisely by Eq.~\eqref{eq:time symmetric},
contributes to wasted heat in any computer. It adds on top of other
beyond-Landauer work bounds including (i) the cost of modular computations,
where global correlations are dissipated due to localized control
\cite{Boyd17a,Riec18} and (ii) the cost of neglecting the local statistics of
the manipulated memory~\cite{Kolc17, Riec18}.

\subsection{(Non)Reciprocity}

We argued that it is common---in practice and, occasionally, out of
necessity---for a computation to be implemented by transforming metastable
memories with time-symmetric driving. We are now in a position to plainly state
the thermodynamic consequences of this type of computation.

Generically, a physical computation produces and is characterized by the set of
memory transition probabilities: $\{ p(m \to m') \}_{m, m' \in \MSet}$. From
Eq.~\eqref{eq:WorkRequired}, we see that the work required for such a
computation, beyond the change in local free energy (which is often constructed
to be zero), is determined by the \emph{nonreciprocity} of memory transitions.
We say that a memory transition is \emph{reciprocated} if:
\begin{align}
p(m \to m') = p(m'^\dagger \to m^\dagger) ~.
\label{eq:Reciprocity}
\end{align}
\emph{Nonreciprocity} $\NR$ quantifies the deviation from this:
\begin{align}
\NR (m \to m') \equiv  \ln \frac{p(m \to m')}{p(m'^\dagger \to m^\dagger)} ~.
\label{eq:Nonreciprocity}
\end{align}
Notably, it vanishes when Eq.~\eqref{eq:Reciprocity} is satisfied. Keep in mind
that $m^\dagger$ and $m'^\dagger$ are themselves valid memory states, so
nonreciprocity compares the probability of two different memory transitions.

Appendix~\ref{sec:TransSpecFT} establishes the transition-specific version of
Eq.~\eqref{eq:WorkRequired}. To summarize, we first derive a useful
transition-specific fluctuation theorem:
\begin{align}
& \braket{e^{-\beta W}}_{\Pr_{\drive}(\St_{0:\tau} |  \MSt_\tau = m', \St_0 \sim  \LocalEq[x_0]{m})} \nonumber \\
& \quad = e^{-\beta \Delta F_x^{(\MSt)}} \frac{
 \Pr_{\smallReverse(\drive)} \bigl(  \MSt_\tau = m^\dagger  \big|  \St_0 \sim \boldsymbol{\pi}_{x_\tau^\dagger}^{(m'^\dagger)}   \bigr) } { \Pr_{\drive} \bigl(  \MSt_\tau = m'  \big|  \St_0 \sim \LocalEq[x_0]{m}  \bigr)}
  ,
\end{align}
which applies to systems that start in local equilibrium.
The appearance of the local-equilibrium free energies makes this a useful 
generalization of related past results \cite{Engl13}.
By Jensen's inequality, and assuming time-symmetric control,
this then implies 
that the minimal average work required of a memory transition is:
\begin{align}
W_\text{min}^{t\text{-sym}} & (m \!\to\! m')
= F_{x_0}^{(m')} \!-\! F_{x_0}^{(m)} \!+\! \kB T \NR (m \!\to\! m') .
\label{eq:WorkFromNRandLFE}
\end{align}
In the common scenario where all local-equilibrium free energies are constructed to be equal,
this reduces to:
\begin{align}
W_\text{min}^{t\text{-sym}}(m \to m') =  \kB T \NR (m \to m') ~.
\label{eq:WorkFromNR}
\end{align}
In other words, \emph{nonreciprocity implies work}. This can also be seen from
\emph{Trajectory Class Fluctuation Theorems} if one restricts to trajectories
that start and end in the specified memory states \cite{Wims19a}. Not all work
is bad, though. When $W_\text{min}^{t\text{-sym}}(m \to m') $ is negative, some
work can be harvested during the transition. In fact, such work harvesting
is necessary to achieve minimal dissipation.

Equation~\eqref{eq:WorkFromNR} implies that \emph{strictly reciprocal}
computations, for which Eq.~\eqref{eq:Reciprocity} is satisfied for all $m, m'
\in \MSet$, can be implemented with a time-symmetric protocol with no expended
work at all. The identity map is a trivial example. Two-cycles, like a bit
swap, can also be implemented for free if using a time-symmetric memory
substrate (for which $m^\dagger = m$ for all $m \in \MSet$). However, most
computations require nonreciprocated memory transitions and, so, require work.

While the Landauer work cost corresponds to average state-space compression,
nonreciprocity corresponds to a localized imbalance of memory currents. Even in
the case of heterogeneous local free energies (i.e.,
Eq.~\eqref{eq:WorkFromNRandLFE}), the minimal dissipation required of
time-symmetrically driven computations is the difference between the two:
\begin{align}
\langle W_\text{diss} \rangle^{t\text{-sym}}_\text{min} 
  \!=\! \kB T \braket{\NR (\MSt_0 \!\to\! \MSt_\tau)}
  \!-\!\langle W \rangle^\text{Landauer}_\text{min}
  .
\label{eq:TimeSymwNR}
\end{align}

The expected nonreciprocity $\braket{\NR (\MSt_0 \to \MSt_\tau)}$ weights how
often a nonreciprocated transition is exercised. This uses the \emph{weighted
nonreciprocity} $d(m \to m')$:
\begin{align}
d(m \to m') = p(m \to m') \NR(m \to m') ~,
\end{align}
which appeared previously in Eq.~\eqref{eq:dmmSimplified}.
The nonreciprocity weight is critical to the following analysis. Though $\NR(m
\to m') = - \NR(m'^\dagger \to m^\dagger)$, weighted nonreciprocity can
strongly break the symmetry.

Consider a typical computation that enforces logical transitions: some
transitions should happen with certainty while others should be forbidden.  In
particular, consider a nonreciprocated $m \mapsto m'$ logical transition: $m
\mapsto m'$ should be strongly enforced ($p(m \to m') \approx 1$), while
$m'^\dagger \not\mapsto m^\dagger$ should be strongly forbidden ($p(m'^\dagger
\to m^\dagger) < \epsilon \ll 1$ for some error tolerance $\epsilon \to 0$).
Then $d(m \to m')$ diverges as $\ln(1/\epsilon)$ while $d(m'^\dagger \to
m^\dagger)$ vanishes. In short, divergent work is required to implement the
desired nonreciprocated $m \mapsto m'$ logical transition.

A \emph{deterministic} computation $\mathcal{C}$ entails a deterministic
logical transformation $m \mapsto \mathcal{C}(m)$ satisfying the reciprocity
condition of Eq.~\eqref{eq:Reciprocity} only when:
\begin{align}
\mathcal{C} \bigl( \mathcal{C} (m)^\dagger \bigr)^\dagger = m ~.
\label{eq:DeterministicReciprocity}
\end{align}
If Eq.~\eqref{eq:DeterministicReciprocity} is \emph{not} satisfied---i.e., if
$\mathcal{C}(m) = m'$ but $\mathcal{C}(m'^\dagger) \neq m^\dagger$---then the
logical transformation appears to require \emph{infinite} work and
\emph{infinite} dissipation since, from Eq.~\eqref{eq:WorkRequired}, it appears
to require work of $\kB T \ln \frac{1}{0}$. In practice, since dissipation is
bounded, such logical transformations can only be approximated. Reliable logic
(with a probability of error less than some small $\epsilon$) requires
significant dissipation of at least $\kB T \ln (1/\epsilon)$ for each violation
of Eq.~\eqref{eq:DeterministicReciprocity}. Section \ref{sec:NearlyDetComp}
develops the consequences of this new relationship between reliability and
dissipation. That then allows us to explore how the two can be balanced in
several examples.

\subsection{Time Symmetric Memory}

The following explores the implications of these results for
time-reversal-invariant memories---$m = m^\dagger$---in which information is
stored via time-symmetric variables, such as spatial location or physical
conformation. In these cases, Eq.~\eqref{eq:time symmetric} simplifies to:
\begin{align}
\beta & \langle W_\text{diss} \rangle^{t\text{-sym}}_\text{min}
  = \Delta H(\MSt_t) \nonumber \\
  &\quad\qquad + \sum_{m,m'}\actual[0](m)
  p(m \rightarrow m') \ln \frac{p(m \rightarrow m')}{p(m' \rightarrow m)}
  ~.
\label{eq:TimeSymAndMemSym}
\end{align}
As before, minimal dissipation depends directly on the nonreciprocity between
memory transitions, but now we have the simplification that:
\begin{align*}
\NR (m \to m') =  \ln \frac{p(m \to m')}{p(m' \to m)}
  ~.
\end{align*}
And here, reciprocity has an especially straightforward interpretation: If $m$
maps to $m'$, then $m'$ should map back to $m$ with the same probability.
Otherwise work is required and, if it is above the Landauer bound, it must be
dissipated.

What is reciprocity in the limit of deterministic computation? For
time-symmetric memory, it acts somewhat like logical invertibility. However, it
is stricter.  Rather, it is logical self-invertibility. Logical
noninvertibility, in contrast, gives rise to state-space compression---the
origin of the Landauer bound. Indeed, circumventing the latter led early
researchers to investigate \emph{reversible} computing---computing with
invertible logical transformations~\cite{Benn82}. Logically reversible
computing requires no work to operate. However, it is now understood that even
logically irreversible computing can be accomplished with zero dissipation,
despite requiring some recyclable work~\cite{Maro09,Saga14,Parr15a}.

The next section draws out the consequences of nonreciprocity in the limit of
nearly deterministic computation. In contrast to the Landauer cost of logical
irreversibility---which, in the limit of zero error, saturates to some small
value on the order of the thermal energy---the thermodynamic cost of
nonreciprocity \emph{diverges} in the limit of zero error, with no chance of
recovering the input energy. It demands an accounting.

\section{Almost-Deterministic Computing}
\label{sec:NearlyDetComp}

At this point, we showed that the transition probabilities entailed by a
desired logical transformation set a lower bound on the work required to
instantiate it physically. We noted that this gives a much stronger bound than
that due to Landauer, which only depends on the relative state-space volume
supporting memory before and after a computation. More specifically, when
time-symmetric control implements a transformation of metastable memories, the
minimal work investment is proportional to the nonreciprocity (non-self-invertibility)
of memory transitions. Furthermore, we discovered that the time-reversal
symmetries of the memory elements themselves can substantially alter the
minimal thermodynamic costs.

Drawing out the practical consequences, the following shows that computing with minimum work depends directly on a computation's error rate $\epsilon$. Most concretely, time-symmetry-induced dissipation diverges as $\ln (1/\epsilon)$ for almost-deterministic computation, where
metastable memories are transformed at low error rate $\epsilon \to 0$.

Consider the physical memory apparatus of a computer that approximately
implements a deterministic computation $\mathcal{C} \colon \MSet \to \MSet$,
mapping memory states to memory states. Then, the memory-state transition
probabilities appearing in Eq.~\eqref{eq:time symmetric} are strongly
constrained by the computation's desired reliability. That is, the probability
that the nonequilibrium thermodynamic information processing takes a memory
state $m$ to anywhere besides $\mathcal{C} (m)$ must be no greater than the
error tolerance $\epsilon$, with $0 < \epsilon \ll 1$. 
Reliability requires choosing a time-symmetric drive protocol $\drive$ that
guarantees that $\Pr_{\drive} \bigl( \MSt_\tau=m'| \MSt_0=m)  \leq \epsilon$
for all $m, m' \in \MSet$ such that $m' \neq \mathcal{C}(m)$. The thermodynamic
implications for a reliable computation then follow from Eq.~\eqref{eq:time
symmetric} with:
\begin{align*}
p(m \rightarrow m')
\begin{cases}
\geq 1-\epsilon & \text{if } \mathcal{C}(m) = m' \\
\leq \epsilon & \text{if } \mathcal{C}(m) \neq m'
\end{cases}.
\end{align*} 

Evaluating Eq.~\eqref{eq:time symmetric} requires addressing four cases,
depending on whether $\mathcal{C}(m) = m'$ or $\mathcal{C}(m) \neq m'$ and on
whether $\mathcal{C}(m'^\dagger) = m^\dagger$ or $\mathcal{C}(m'^\dagger) \neq
m^\dagger$. Any given implementation results in an \emph{actual} probability of
error for each of the intended transitions: $\epsilon_{m} = 1 -p\big(m
\rightarrow \mathcal C (m)\big)$. We adopt the design constraint that
$\epsilon_{m} \leq \epsilon$ for all possible initial memories $m$. Then, given
an implementation $\drive$, the probability of an \emph{accidental} memory
transition is $\epsilon_{m \rightarrow m'} =p(m \rightarrow m')$ for $m' \neq
\mathcal{C}(m)$. Since $\sum_{m' \in \MSet \setminus \{ \mathcal{C}(m) \} }
\epsilon_{m \rightarrow m'} = \epsilon_{m} \leq \epsilon$, we must have that $0
\leq \epsilon_{m \to m'} \leq \epsilon_{m} \leq \epsilon$.

To simplify, we temporarily restrict to the case of time-reversible memories,
where $m=m^\dagger$. 
The four cases then depend
on whether $\mathcal{C}(m) = m'$ or
$\mathcal{C}(m) \neq m'$, and on whether $\mathcal{C}(m') = m$ or
$\mathcal{C}(m') \neq m$. Later, we quote the general result that does not
require $m=m^\dagger$.

Appendix~\ref{sec:4cases} uses the error constraints to evaluate
the weighted nonreciprocity:
\begin{align*}
d(m,m')=p(m \rightarrow m') \ln \frac{p(m \rightarrow m')}{p(m' \rightarrow m)}
  ~,
\end{align*}
for the four possible cases, finding:\\[-20pt]
\begin{enumerate}
\setlength{\topsep}{-15pt}
\setlength{\itemsep}{-15pt}
\setlength{\parsep}{-15pt}
\item $\mathcal{C}(m) = m'$; $\mathcal{C}(m') = m$:\\[-20pt]
\begin{align*}
- \epsilon \leq d^{(1)}(m,m')
  \leq \epsilon + \tfrac{1}{2} \epsilon^2 + \mathcal{O}(\epsilon^3)
  ~.
\end{align*}
\item $\mathcal{C}(m) = m'$; $\mathcal{C}(m') \neq m$:\\[-20pt]
\begin{align*}
\ln(\epsilon^{-1}) \lesssim d^{(2)}(m,m') \leq  \ln (\epsilon_{m' \to m}^{-1})  
  ~.
\end{align*}
\item $\mathcal{C}(m) \neq m'$; $\mathcal{C}(m') = m$:\\[-20pt]
\begin{align*}
\epsilon_{m \to m'} \ln \epsilon_{m \to m'} < d^{(3)}(m,m') < 0
  ~.
\end{align*}
\item $\mathcal{C}(m) \neq m'$; $\mathcal{C}(m') \neq m$:\\[-20pt]
\begin{align*}
-\epsilon/e \leq d^{(4)}(m,m') = \epsilon_{m \to m'}
  \ln \left( \frac{\epsilon_{m \to m'}}{\epsilon_{m' \to m}} \right)
  ~.
\end{align*}
\end{enumerate}
In the high-reliability limit $\epsilon \to 0$, most $d^{(n)}(m,m')$ terms
are on the order of $\epsilon$ and so tend to zero. That is, \emph{except} for
$d^{(2)}(m,m')$ in Case 2 whose contribution to the dissipation \emph{diverges}
as $\epsilon \to 0$ since $d^{(2)}(m,m') \gtrsim \ln(\epsilon^{-1})$.

Case 4 is somewhat more delicate than Cases 1 and 3, due to the ratio of
errors. When $\epsilon_{m' \rightarrow m} \geq \epsilon_{m \rightarrow m'}$,
then $- \epsilon / e \leq d^{(4)}(m,m') \leq 0$, so that $d^{(4)}(m,m')$
vanishes as $\epsilon \to 0$. However, when $\epsilon_{m \rightarrow m'} >
\epsilon_{m' \rightarrow m}$, there is a chance for slightly more dissipation.
Nevertheless, $d^{(4)}(m,m')$ is still only on the order of $\epsilon$, so long
as the relative error rates $\epsilon_{m \rightarrow m'}$ and $\epsilon_{m'
\rightarrow m}$ are within several orders of magnitude of each other. If
$\epsilon_{m \rightarrow m'} \gg \epsilon_{m' \rightarrow m}$, then extra
dissipation will be incurred due to the stringent reliability of the forbidden
$m \nrightarrow m'$ transition, beyond the design constraint.

The upper bound in Case 2 also implies extra dissipation when $\epsilon_{m' \to
m}$ is more reliable than the reliability design constraint. And, this
extra dissipation need not be small. Generically, though, for a given
error tolerance, we expect that minimal dissipation can be achieved by allowing
all error rates to be as close to uniform as possible, meeting but not
significantly exceeding the overall reliability constraint. Still, due to Case
2, even the minimal dissipation diverges with increasing reliability.

Thus, Case 2 provides the only contribution of the $d(m, m')$ terms to
Eq.~\eqref{eq:WdissDecomp} in the low-error limit. Since the target computation
is deterministic---$\mathcal{C}$ maps each memory state $m$ to one memory state
$\mathcal C (m)$---there can be at most one contribution of $d^{(2)}(m, m')$
for each $m$, coming from $m' = \mathcal C (m)$, and only if $\mathcal C (m')
\neq m$.  The total contribution from each $m$ is:
\begin{align*}
\IverL \mathcal{C}(\mathcal{C}(m) ) \neq m \IverR  \ln (\epsilon_{\mathcal{C}(m) \rightarrow m}^{-1})
  ~,
\end{align*}
where $\IverL \cdot \IverR$ is the Iverson bracket, which returns $1$ when its
argument is true and $0$ otherwise. In this case,
$\IverL \mathcal{C}(\mathcal{C}(m) ) \neq m \IverR = 
1 -   \delta_{m, \mathcal{C}(\mathcal{C}(m) ) } $.

Applying the low-$\epsilon$ contributions to Eq.~\eqref{eq:dmmSimplified}
yields an approximate bound for the time-symmetric work. If the memories have
the same local-equilibrium free energy, the bound is:
\begin{align}
\beta \braket{W}^{t\text{-sym}}_\text{min}
  & \approx \sum_{m \in \MSet} \actual(m)
  \IverL  \mathcal{C}(\mathcal{C}(m) ) \neq m  \IverR
  \ln (\epsilon_{\mathcal{C}(m) \rightarrow m}^{-1}) \nonumber \\
  & \geq  \ln (\epsilon^{-1}) \!\!\! \sum_{m \in \MSet} \actual(m)
  \IverL  \mathcal{C}(\mathcal{C}(m) ) \neq m  \IverR   \label{eq:Wtsym}  \\
  & = \beta \langle W \rangle_\text{min}^\text{approx}
  ~.
\nonumber
\end{align}
This determines how the tradeoff between work and reliability scales for
almost-perfect computation. And, it identifies the role that a computation's
reciprocity plays in dissipation.

The result is a simple error--dissipation tradeoff, if we restrict ourselves
to using time-reversible protocols for implementing reliable computations:
\begin{align}
\beta \braket{\Wdiss}_\text{min}^{t\text{-sym}} 
  & \gtrsim \beta \langle \Wdiss \rangle_\text{min}^\text{approx} \nonumber \\
  & = \bigl\langle \IverL \mathcal{C}(\mathcal{C}(\MSt_0) )
  \neq \MSt_0  \IverR \bigr\rangle_{\MSt_0} \ln(\epsilon^{-1} ) \nonumber \\
  & \qquad + \Delta H(\MSt_t)
  ~.
\label{eq:MainScalingResult}
\end{align}
Since the change $\Delta \Shannon( \MSt_t)$ in entropy is finite as $\epsilon
\to 0$, the divergent nonreciprocity contribution to the dissipation dominates.
Moreover, the simple scaling of the reliability--dissipation tradeoff is the
same as that for the reliability--work tradeoff. In the low-error limit, 
the minimal nonreciprocity depends only on the computation:
\begin{align*}
\braket{\NR (\MSt_0 \to \MSt_\tau)} \geq
\bigl\langle \IverL \mathcal{C}(\mathcal{C}(\MSt_0) )
  \neq \MSt_0  \IverR \bigr\rangle_{\MSt_0} \ln(\epsilon^{-1} ) .
\end{align*}
The reciprocity coefficient:
\begin{align*}
\bigl\langle \IverL \mathcal{C}(\mathcal{C}(\MSt_0) ) \neq \MSt_0 \IverR
  \bigr\rangle_{\MSt_0} 
  & = \sum_{m \in \MSet} \actual(m)
  \IverL  \mathcal{C}(\mathcal{C}(m) ) \neq m  \IverR 
\end{align*}
is the probability that the memory makes a transition that is \emph{not}
reciprocated if the output becomes the input to the computation. In short,
\emph{computations with nonreciprocity require significant dissipation when
implemented with a time-symmetric protocol}.

A sequel explores the dynamical mechanism for the dissipation in this case.
With time-symmetric protocols, state-space compression occurs only via a loss
of stability and subsequent dissipation. This corresponds to a topological
restriction on the bifurcation structure of local attractors---a restriction
imposed by time-symmetric control.

\begin{figure}[H]
\includegraphics[width=0.7\columnwidth]{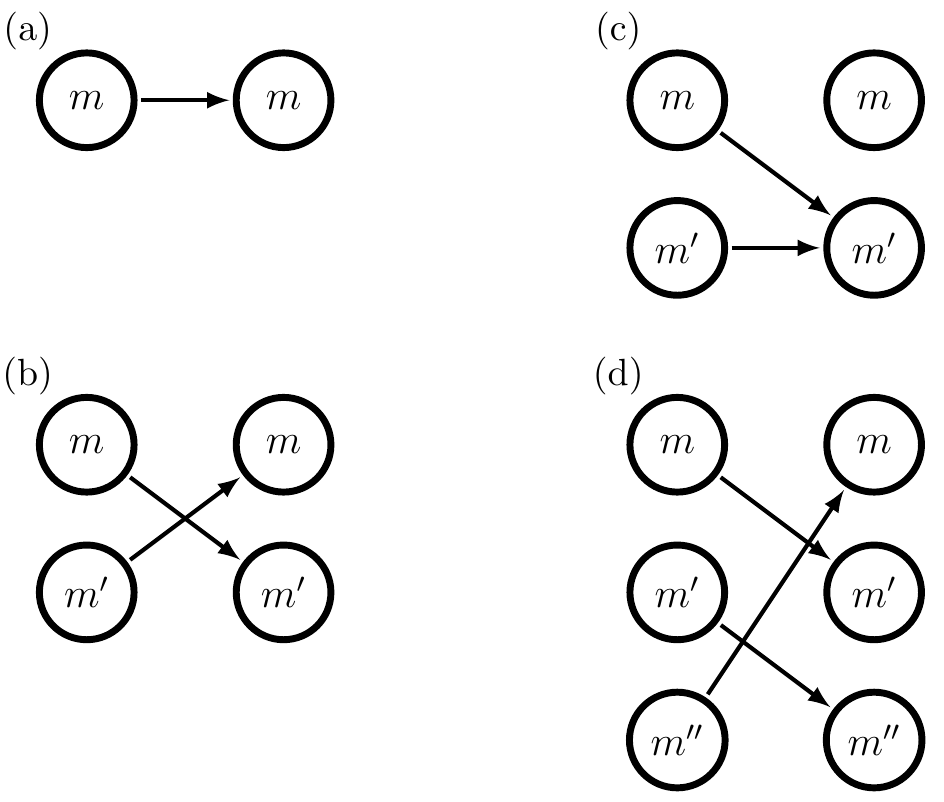}
\centering
\caption{Simple features of deterministic computations determine the minimal
	dissipation required to implement them reliably with time-symmetric
	protocols. Dissipation is incurred for nonreciprocated memory transitions.
	When using time-reversal-invariant memory elements ($m^\dagger = m$): (a)
	identity mappings and (b) swap operations can be implemented freely,
	whereas (c) memory-space compression and (d) nonreciprocated permutations
	require dissipation that diverges with increasing reliability.
	}
\label{fig:ReciprocalAndNonreciprocalChannels} 
\end{figure}

One also concludes that strictly-reciprocal logic gates, such as the identity
(communication relay) and NOT gates, where $\mathcal{C}(\mathcal{C}(m))$ always
returns $m$, are \emph{exempt} from this dissipation. One can efficiently
implement them with time-symmetric protocols. For all other computations,
time-\emph{asymmetric} control must be used to avoid this dissipation.
Figure~\ref{fig:ReciprocalAndNonreciprocalChannels} gives examples of
reciprocated and nonreciprocated memory transitions.  

The main result Eq.~\eqref{eq:Wtsym} generalizes to the case where $m, \,
m^\dagger \in \MSet$ without the requirement that $m=m^\dagger$. Then,
dissipation diverges with increasing reliability whenever $m \to m'$
transitions are made such that $\mathcal{C}(m) = m'$ but
$\mathcal{C}(m'^\dagger) \neq m^\dagger$. For uniform local free energies,
Eq.~\eqref{eq:Wtsym} becomes:
\begin{align}
\beta \braket{W}^\text{approx}_\text{min} 
  = \ln (\epsilon^{-1}) \!\!\!\! \sum_{m \in \MSet} \!\!\!\! \actual(m) \IverL  \mathcal{C}(\mathcal{C}(m)^\dagger ) \neq m^\dagger  \IverR.
\label{eq:Wdisstsym_masym}
\end{align}

Notably, this allows one to quantify error--work and error--dissipation
tradeoffs when computing with magnetic memory systems, for example. For a
single bistable magnetic device, $m$ could represent the $\one$ memory of
having an ``up'' magnetic moment or ``clockwise'' in the case of toroidal
magnetic core memory elements. And, $m^\dagger$ would then represent the
``down'' or ``counter-clockwise'' memory $\zero$, as shown on the right side of
Fig. \ref{fig:ConjugateMemory}. This implies that memories with certain types of
symmetry are better suited for certain types of logical operation, in that they
minimize nonreciprocity and so dissipation.

To emphasize, the dissipations here are distinct from the heat associated with
logically-irreversible transformations, as discussed by Landauer and
Bennett~\cite{Land61a, Benn82}, which arises as a compensation to microscopic
state-space contraction. There are two important distinctions.

First, the heat devolved with logical irreversibility---e.g., the minimal heat
of $\kB T \ln 2$ released to the environment upon erasure---is \emph{not
necessarily irreversibly dissipated}~\cite{Maro09, Saga14, Parr15a, Riec18}.
It is offset by a change in the nonequilibrium addition to free energy and so
can be leveraged later to do an equal amount of useful work.
In contrast, nonreciprocity dissipation is
energy that is truly dissipated. It is irretrievably lost to the environment,
with no compensation via change of the nonequilibrium addition to free energy
and so can never be recovered.

Second, logical nonreciprocity is distinct from logical irreversibility.
Strictly-reciprocal computations, where $\mathcal{C}(\mathcal{C}(m)) = m$ for
all $m \in \MSet$, are logically reversible, being their own inverses.
However, logically reversible permutations of the memory can be completely
nonreciprocal, as demonstrated in
Fig.~\ref{fig:ReciprocalAndNonreciprocalChannels}(d). Reciprocity requires not
only that the deterministic logic be invertible, but further that the logical
dynamic inverts itself. In short, the Landauer work cost corresponds to logical
noninvertibility, while nonreciprocity cost corresponds to logical
\emph{noninvolution}. The reciprocity bound on dissipation can therefore be
interpreted as the minimal work required to implement a reliable computation
with time-symmetric protocols \emph{in addition to} the well-known Landauer
bound.

\section{Applications}

With the basics of time-symmetric and nonreciprocal computing in hand, we turn
to explore the thermodynamic implications for erasure, logic gates, and
biological information processing. Our companion work Ref. \cite{Wims20a} gives
a detailed analysis of erasure and explores several model implementations.

\subsection{Erasure}

To directly illustrate how the cost of time-symmetric control differs from
Landauer's bound, consider the classic example of bit erasure. Landauer
originally described a method for erasing a bit of information stored in a
double-well potential in contact with a heat bath: tilt the potential---moving
overdamped stochastic particles to one side---and return the potential to its
original orientation, leaving the particles (temporarily) trapped in one well
\cite{Land61a}. If implemented naively, via a protocol that raises the energy
of one well and then lowers it at the same rate, the time-symmetry results in
dissipation significantly above Landauer's bound, as described by the bounds
just developed. Note that protocol time-symmetry does \emph{not} imply spatial
symmetry---one can tilt the potential to the left without needing to tilt it to
the right.

\begin{figure}[h]
\includegraphics[width=.5\columnwidth]{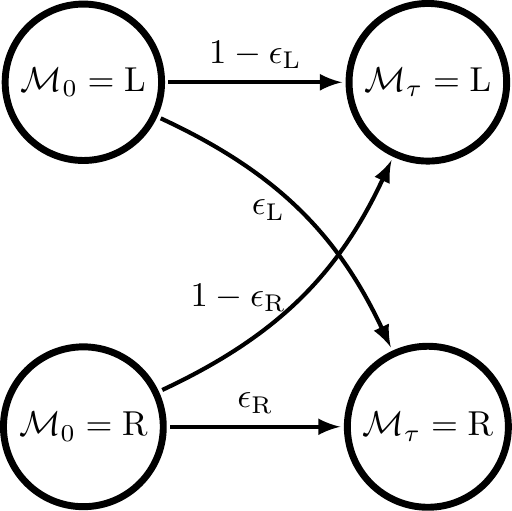}
\centering
\caption{Markov transition matrix for any implementation of erasure gives
	the error rate $\epsilon_\Left=\epsilon_{\Left \rightarrow \Right}$ from
	initial memory state $\mathcal{M}_0=\Left$ and the error rate
	$\epsilon_\Right=\epsilon_{\Right \rightarrow \Left}$ from the initial
	memory state $\mathcal{M}_0=\Right$.
	}
\label{fig:ErasureChannel} 
\end{figure}

Bit erasure $\mathcal{C}_\text{erase}$ operates on a single bistable element
with memory states $\MSet = \{ \Left, \Right \}$ that correspond to occupying
the Left or Right side of a double-well potential energy landscape. The
computation is defined by resetting to the $\Left$ memory state:
$\mathcal{C}_\text{erase}(\Left) = \Left$ and $\mathcal{C}_\text{erase}(\Right)
= \Left$. Hence, $\mathcal{C}_\text{erase}( \mathcal{C}_\text{erase} (\Left)) =
\Left$, while $\mathcal{C}_\text{erase}(\mathcal{C}_\text{erase}(\Right)) \neq
\Right$. Thus, the net transition probabilities $p(m \rightarrow m')$ shown in
Fig.~\ref{fig:ErasureChannel} characterize any reliable implementation of this
computation in terms of the probabilities of errors,
$\epsilon_\Right=\epsilon_{\Right \rightarrow \Right}$ and
$\epsilon_\Left=\epsilon_{\Left \rightarrow \Right}$, that leave the system in
the $\Right$ state.

From this, the only terms contributing to the bound on time-symmetric work are:
\begin{align*}
d(\Left, \Right)
  &= \epsilon_\Left \ln \frac{\epsilon_\Left}{1-\epsilon_\Right} ~\text{and}\\
 d(\Right, \Left)
  & = (1-\epsilon_\Right) \ln \frac{1-\epsilon_\Right}{\epsilon_\Left}
  ~.
\end{align*}
Allowing for a potentially nonuniform input distribution $\actual =
(1-p_\Right, \, p_\Right)$ over memory states yields the exact bound on
time-symmetric work investment:
\begin{align}
\beta \braket{W}_\text{min}^{t\text{-sym}}
  & = (1-p_\Right)d(\Left,\Right)+p_\Right d(\Right,\Left) \nonumber \\
  & = (p_\Right (1-\epsilon_\Right)-(1-p_\Right)\epsilon_\Left)
  \ln \frac{1-\epsilon_\Right}{\epsilon_\Left} \nonumber \\
  & = ( p_\Right- \langle \epsilon \rangle)
  \ln \frac{1-\epsilon_\Right}{\epsilon_\Left}
  ~,
\label{eq:ErasureWorkBound}
\end{align}
where $\langle \epsilon \rangle=p_\Right \epsilon_\Right+(1-p_\Right)
\epsilon_\Left$ is the average error of the computation.

\begin{figure}[t]
\includegraphics[width=.96\columnwidth]{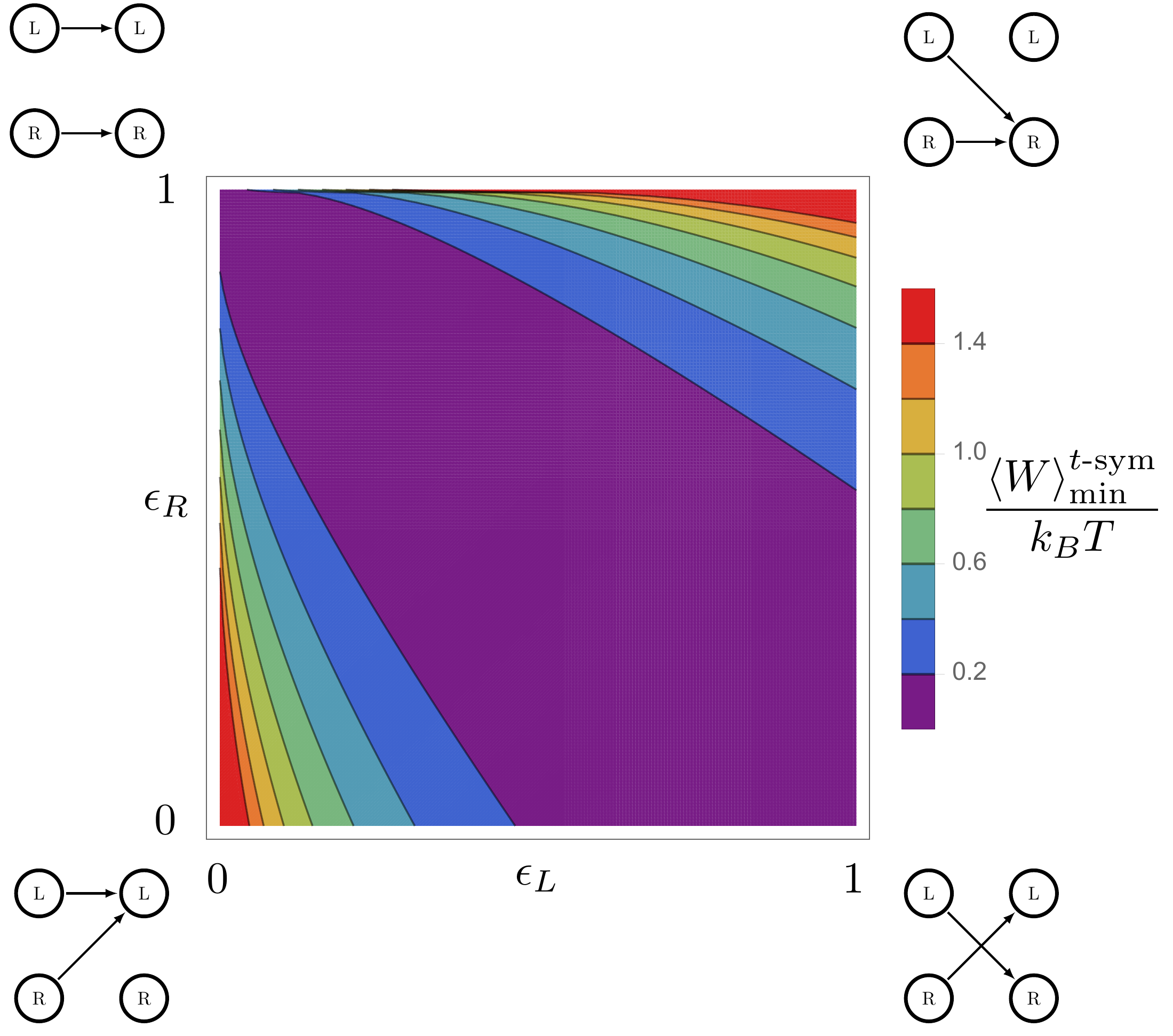}
\centering
\caption{Average minimum work (units of $\kB T$) required for a $1$-bit
	operation diverges for low $\epsilon_\Left$ or high $\epsilon_\Right$,
	corresponding to erasure to $\Left$ or erasure to $\Right$, respectively.
	The state-transition diagrams at the corners show the basic computations
	associated with each extremal point in computation parameter space. The
	identity map (upper left), the bit-flip (lower right), and every operation
	interpolated between require no work according to the time-symmetric
	control bound. 
	The required work diverges at perfect erasure to Left (bottom-left corner) 
	and perfect erasure to Right (top-right corner),
	where the red coloring covers all work values from 1.4 to $\infty$ $\kB T$.
	}
\label{fig:MinimumWork} 
\end{figure}

\begin{figure}[h]
\includegraphics[width=\columnwidth]{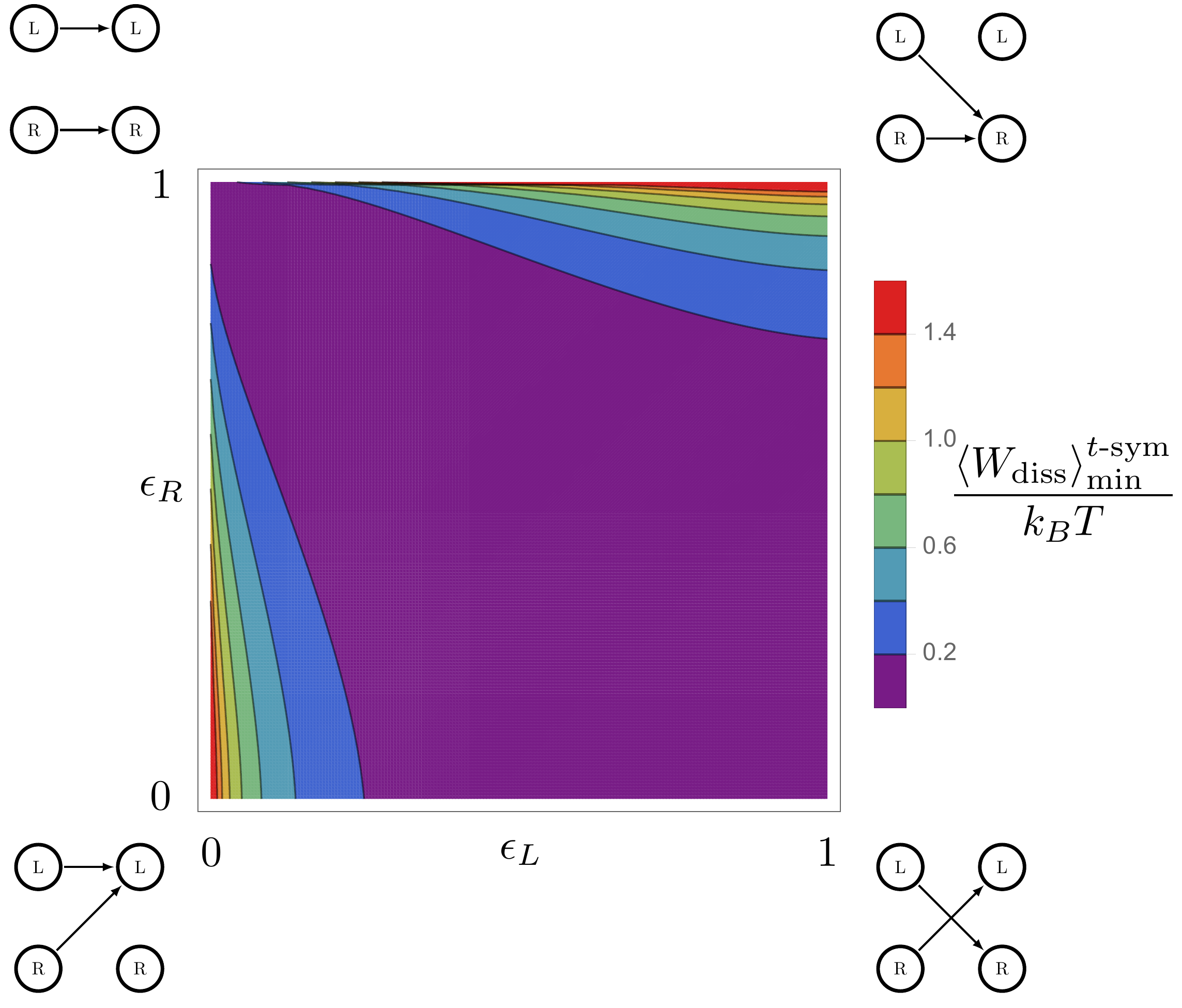}
\centering
\caption{Dissipated work---that required beyond the Landauer bound---behaves
	much the same as the minimum work over possible bit computations, diverging
	for high values of $\epsilon_\Right$ and low values of $\epsilon_\Left$ and
	vanishing for the identity map and bit flip.
	Dissipation diverges at the bottom-left and top-right corners,
	where the red coloring covers all values of dissipated work from 1.4 to $\infty$ $\kB T$.
	}
\label{fig:MinimumDissipation} 
\end{figure}

Figure \ref{fig:MinimumWork} plots the work requirements for an initially
unbiased memory state with $p_\Right=1/2$. We see that the work diverges for
small values of error $\epsilon_\Left$. The plot also indicates that the work
diverges for high values of error $\epsilon_\Right$, due to the symmetry of the
system between left and right. It is also worth noting from the plot that there
are computations, such as bit flips, which this bound suggests may be
achievable with time-symmetric control and without energetic cost. However, we
are primarily concerned with the lower left portion of the plot, where
effective erasures occur. The divergent scaling of the work required to
reliably erase overwhelms the rather meager energy requirements given by
Landauer's bound. And, unlike the latter, this excess work requirement must
be irreversibly dissipated. Subtracting off the Landauer bound $\kB T \Delta
H(\MSt_t)$ to calculate the minimum dissipation $\langle
W_\text{diss}\rangle_\text{min}^{t\text{-sym}}$,
Fig.~\ref{fig:MinimumDissipation} shows that the divergent contribution to the
work is attributed to nonreciprocity.

Applying Eq.~\eqref{eq:Wtsym}'s work bounds to erasure, we see that:
\begin{align*}
\bigl\langle \IverL \mathcal{C}_\text{erase}(\mathcal{C}_\text{erase}(\MSt_0)
)
  \neq \MSt_0  \IverR \bigr\rangle_{\MSt_0} = p_\Right
  ~.
\end{align*}
This means that, in the low-error limit:
\begin{align}
\langle W\rangle^\text{approx}_\text{min}
  = p_\Right  \ln(\epsilon^{-1} ) \, \kB T
  ~.
\label{eq:SymErasureScaling}
\end{align}
The initial memory-state entropy is $H_\text{b} ( p_\Right ) \equiv - p_\Right
\ln p_\Right - (1- p_\Right) \ln (1- p_\Right) \leq \ln 2$, which is called the
\emph{binary entropy function} \cite{Cove06a}. Whereas, the final memory-state
entropy vanishes as $\epsilon \to 0$.  Hence, from
Eq.~\eqref{eq:MainScalingResult}, we immediately find that the small-error
dissipation necessary for time-symmetric erasure diverges as $\epsilon \to 0$:
\begin{align}
\braket{\Wdiss}^{\text{approx}}_\text{min}
  \approx p_\Right  \ln(\epsilon^{-1} ) \, \kB T
  - H_\text{b} ( p_\Right ) \, \kB T
  ~.
\label{eq:SymErasureScalingResult}
\end{align}
Landauer erasure typically assumes a uniform initial distribution where
$p_\Right = 1/2$. This results in $\braket{\Wdiss}^{t\text{-sym}} \gtrsim
\tfrac{1}{2} \ln(1 / 4 \epsilon ) \, \kB T$. Since the memory-state entropy is
bounded by $\kB T \ln 2$, the contribution from Landauer's bound is negligible
compared to the $\ln (\epsilon^{-1})$ term for small $\epsilon$. And so, the
latter dominates both the work and dissipation for high-fidelity erasure. 

To test the strength of these bounds and their approximations,
Ref.~\cite{Wims20a} considers two different implementations of time-symmetric
erasure---rate equations for a two-level system and one-dimensional Langevin
dynamics in a controlled double-well potential.  The example implementations
not only validate our bounds, but moreover show that the bounds are tight.

\subsection{Logic Gates}
\label{sec:LogicGates}

We next consider the minimal dissipation required to implement conventional
two-input one-output logic gates that serve as the basis for modern digital
computing: AND, NAND, OR, NOR, XOR, and the like. The typical implementation
requires two bistable memory elements ($\MSt_t^{(\text{in}_1)}$ and
$\MSt_t^{(\text{in}_2)}$) for the input and another bistable memory element
($\MSt_t^{(\text{out})}$) to robustly store the output. Each memory element can
take on one of two memory states $\MSt_t^{( \, \cdot \, )} \in \{ \zero, \one
\}$. Each logic gate is a computation $\mathcal{C}$ on the set of composite
memory states $\MSet =\MSet^{(\text{in}_1)}\times \MSet^{(\text{in}_2)} \times
\MSet^{(\text{out})}= \{\zero \zero \zero,\zero \zero \one, \zero \one \zero,
\dots \one \one \one \}$.

In particular, since networks of NAND gates are sufficient for universal
computation, the NAND gate is worthy of immediate investigation.
Appendix~\ref{sec:NAND} analyzes the NAND gate, showing that: \emph{if the
memory elements are initiated statistically independent of each other}, then
the minimal time-symmetric-implementation dissipation is comparable to that
expected for bit erasure of the output bit, depending on how often the output
bit is overwritten. However, when the memory system is initialized with
correlation among the memory elements (perhaps as the correlated output of
previous upstream computations), then the change in memory entropy is $\Delta
\Shannon (\MSt_t) \approx - \Shannon \bigl( \MSt_0^{(\text{out})} \big|
\MSt_0^{(\text{in}_1)} , \MSt_0^{(\text{in}_2)} \bigr)$ in the small $\epsilon$
limit, and the reciprocity coefficient is: 
\begin{align}
& \bigl\langle \IverL
\mathcal{C}_\text{NAND}(\mathcal{C}_\text{NAND}(\MSt_0) ) \neq \MSt_0  \IverR
\bigr\rangle_{\MSt_0}  \nonumber \\
& \qquad = \actual(\zero \zero \zero) + \actual(\zero \one \zero)
+ \actual(\one \zero \zero) + \actual(\one \one \one) ~. 
\end{align}
As a consequence, the
initial correlation among the memory elements has a profound impact on the
coefficient of minimal dissipation. The overall dissipation still diverges
$\sim \ln( \epsilon^{-1})$, though, with increasing reliability as long as the
implementation is time-symmetric.

This example highlights a previously unsuspected source of unnecessary and
preventable dissipation in digital computers. Indeed, computations are currently
implemented in hardware, whether via periodic clocking or static gate voltages,
through time-symmetric driving. Error rates in today's reliable digital
computers are extremely low. The so-called \emph{soft error rate} (also known
as \emph{single-event upsets}) for $50$ nm gate technology has been estimated
to be on the order of $10^{-4}$ errors per $10^9$ hours of operation of a CMOS
logic circuit~\cite{Shiv02}. Moreover, these errors are overwhelmingly due to
cosmic rays rather than conventional thermal fluctuations. If the computer is
properly shielded, the error rate is significantly lower. Using the
aforementioned numbers for a conservative estimate and assuming a 3 GHz clock
rate, this extreme reliability of $\epsilon \lesssim 10^{-26}
~\text{errors}/\text{instruction}$ would imply at least 30 $\kB T$ is
dissipated in nearly every elementary operation in modern circuitry directly
due to time-symmetric implementation. This is well beyond the Landauer bound
that, for NAND and any other traditional two-input--one-output logic gate, is
given by $-\Delta H(\MSt_t) \approx \Shannon \bigl( \MSt_0^{(\text{out})} \big|
\MSt_0^{(\text{in}_1)} , \MSt_0^{(\text{in}_2)} \bigr) \leq \ln 2$.

By way of contrast, the universal Fredkin gate---swap the state of the first
two memory elements only if the third memory element is in state $\one$---is
not only logically reversible, but is also a strictly-reciprocal gate
\cite{Fred82b}. As a strictly-reciprocal gate, it can implement universal
computation efficiently using time-symmetric control. This makes Fredkin gates
an especially tempting basis for energetically-efficient future computing
technologies, assuming they can be made to transform metastable memories
without the exponential sensitivity to initial conditions that condemned their
initially-proposed mechanical instantiation~\cite{Fred02,Benn82}.

\subsection{Dissipation in Biological Processing}
\label{sec:NESSs}

Beyond logical computation, the reliability-energy-efficiency tradeoff also
applies to information processing in far-from-equilibrium biological systems.
For this, it is useful to reformulate Eq.~\eqref{eq:WdissGeneralMicro} to apply
when $\Wdiss$ is the work dissipated in controlling (through $\drive$) a system
already maintained out of equilibrium by a given housekeeping entropy
production $\Sigma_\text{hk}$. This can happen, for example, via ATP hydrolysis
or the maintenance of a chemical potential gradient. Then, according to
Eq.~(27) of Ref.~\cite{Riec17}, the more general starting point is:
\begin{align}
\beta & \Wdiss + \kB^{-1} \Sigma_\text{hk} \nonumber \\
  & = \ln \left(  
\frac{\Pr_{\drive} \bigl( \St_{0:\tau} = \sseq | \St_0 \sim \actual[0] \bigr) }{\Pr_{\smallReverse(\drive)} \bigl( \St_{0:\tau} = \Reverse(\sseq) | \St_0 \sim \actual[\tau]^\dagger \bigr)}
\right) ~.
\label{eq:WdissShkGeneralMicro}
\end{align}
In steady-state biological processes, there is no external control. The controllable parameters can thus be considered as held constant, such that the driving protocol is trivially time-symmetric $\Reverse(\drive) = \drive = x_0 x_0 \dots x_0$ and $\Wdiss = 0$. That is, the system is driven solely by the housekeeping entropy production---power supplied from a stationary stochastic influence---that maintains the steady state. For example, molecular machines can sustain time-asymmetric functionality within their time-symmetric ambient chemical environment\footnote{This time-symmetric ambient chemical environment
is maintained by power-consuming homeostatic mechanisms elsewhere in a cell.}
due to the entropy they produce during chemical catalysis~\cite{Feng08, Astu17}.
In these nonequilibrium steady states, the expected microstate distribution of
the system is the time-invariant steady-state distribution:
\begin{align}
\actual[0] = \actual[\tau] = \stationary_\text{s.s.} 
\end{align}
Moreover, due to the low Reynolds numbers of the microbiological realm,
momentum is heavily damped and the relevant memory states are
time-reversal-invariant: $m^\dagger = m$ for all $m \in \MSet$. We also
have $\stationary_\text{s.s.}^\dagger = \stationary_\text{s.s.}$.

Integrating these generic features of steady-state biological transformations,
Eq.~\eqref{eq:WdissShkGeneralMicro} simplifies to:
\begin{align}
\kB^{-1} \Sigma_\text{hk} 
  \! = \! \ln \!\left(  \!
\frac{\Pr \bigl( \St_{0:\tau} \!=\! \sseq  | \St_0 \!=\! s_0 \bigr)
\stationary_\text{s.s.} (s_0) }{\Pr \bigl( \St_{0:\tau} \!=\! \Reverse(\sseq) |
\St_0 \!=\! s_\tau \bigr)  \stationary_\text{s.s.}(s_\tau)}
\! \right) \!.
\label{eq:ShkMicro}
\end{align}
The single-timestep version of this is well known:
\begin{align*}
\delta \Sigma_\text{hk} 
  & = \kB  \ln \left(  
\frac{\Pr \bigl( \St_{\delta t} = s_{\delta t}  | \St_0 = s_0 \bigr)  \stationary_\text{s.s.} (s_0) }{\Pr \bigl( \St_{0:\tau} = s_0| \St_0 = s_{\delta t} \bigr)  \stationary_\text{s.s.}(s_{\delta t})}
\right) ~.
\end{align*}
As $\delta t \to 0$, the relative probabilities reduce to the relative transition rates $\{ r_{s \to s'} \}$:
\begin{align}
\delta \Sigma_\text{hk} 
  & = \kB  \ln \left(  
\frac{  \stationary_\text{s.s.}  (s_0) \, r_{s_0 \to s_{\delta t} }  }{ \stationary_\text{s.s.}(s_{\delta t}) \,  r_{s_{\delta t} \to s_0 }  }
\right) ~.
\label{eq:ShkMicroSingleStep}
\end{align}
(Metastability of mesoscopic conformations allows
Eq.~\eqref{eq:ShkMicroSingleStep} to hold approximately for transitions between
mesoscopic states as well.) For example, Eq.~\eqref{eq:ShkMicroSingleStep} was
the starting point for the recently-derived thermodynamic uncertainty
relations that received attention \cite{Ging16}.

However, much can be learned from approaching finite-duration transformations
directly, by coarse-graining Eq.~\eqref{eq:ShkMicro} according to initial and
final functionally-relevant states. Notably, it allows simple yet powerful
analysis of the minimal dissipation required for biological functionality,
regardless of how complicated the finite-duration biological implementations
appear. And, it implies a generic error--dissipation tradeoff that we expect to
apply broadly to the reliable performance of microbiological systems.

Working this out explicitly leads to a pleasantly simple picture of
steady-state biological information processing and its requisite dissipation:
\begin{widetext}
\begin{align}
\kB^{-1} \braket{ \Sigma_\text{hk} }
& \geq 
  \DKL{\Pr(\MSt_0 = m, \MSt_\tau = m')}{\Pr(\MSt_\tau = m', \MSt_{2
  \tau} = m)} 
  \nonumber \\
&= \underbrace{\Delta \H (\MSt_t) }_{\text{\scalebox{0.8}{$=0$ since $\actual[\tau] = \actual[0] =  \stationary_\text{s.s.}$}}}   
+  \sum_{m, m' \in \MSet}
\stationary_\text{s.s.}(m) \, p(m \to m') \ln \frac{p(m \to m')}{ p(m'
\to m)} 
\nonumber \\
&= 
  \sum_{m, m' \in \MSet}  \stationary_\text{s.s.}(m) \, p(m \to m') \NR (m \to m') 
  \nonumber \\
&= 
  \braket{\NR(\MSt \to \MSt')}_{\stationary_\text{s.s.}(\MSt) \, p(\MSt \to \MSt')}   ~.
\label{eq:SSNR}
\end{align}
\end{widetext}
Steady-state computation quite simply becomes enforcing a Markov model for
memory transitions $\{ p( m \to m' )\}$. The minimal dissipation achievable via
any biological implementation then equals or exceeds the expected nonreciprocity. It
should be noted that the steady-state distribution over memory states
$\stationary_\text{s.s.}$ is easily found as the stationary eigenstate of the
Markov model's transition matrix.

In equilibrium, where $\stationary_\text{s.s.}=\stationary_\text{eq}$, the
average nonreciprocity vanishes since detailed balance demands
$\stationary_\text{eq}(m) p_\text{eq}(m \to m') = \stationary_\text{eq}(m')
p_\text{eq}(m' \to m) $ while, as always, $\NR(m \to m') = - \NR(m' \to m)$.
As necessary, this allows zero dissipation in equilibrium. In contrast,
nonequilibrium steady-state computations must dissipate since they bias
nonreciprocated transitions. These computations enforce nondetailed-balanced
dynamics.

Equation~\eqref{eq:SSNR} shows that dissipation in steady-state biological
processes is bounded by the average nonreciprocity. Remarkably, this bound
depends only on the net transformation and is independent of intermediary
steps and biochemical details.

Thus, as in the preceding analysis, one expects a generic error--dissipation
tradeoff: 
\begin{align*}
d \Sigma_\text{hk} / dt \propto \ln (1/ \epsilon) 
\end{align*}
for reliable steady-state biological transformations of genetic
information---whenever certain transitions need to be strongly enforced with
only a small probability of failure $\epsilon$. The generic error--dissipation
tradeoff  holds, as an important example, for processes such as chemical
proofreading necessary for reliable DNA replication. In this way, our
contribution dovetails with more model-specific results on the latter by
Refs.~\cite{Hopf74, Benn79, Muru12, Sart15}. Since codons are nonreciprocally
corrected to satisfy Watson--Crick base pairing, each correction requires
significant dissipation.

The $d \Sigma_\text{hk} / dt \propto \ln (1/ \epsilon)$ error--dissipation
tradeoff should arise in other reliable biological transformations as well.
For example, to reliably maintain a target nonequilibrium state
Ref.~\cite{Horo17} finds exactly this minimum dissipation rate. This is
relevant to protein folding, as well as other biological functionalities.

Our results also support and update the perspective of Ref.~\cite{Ould17},
where steady-state biochemical copying networks (e.g., for sensing) were found
to have an unavoidable error--dissipation tradeoff due to time-invariant
driving. Our results more broadly suggest error--dissipation tradeoffs for any
biochemical task with nonreciprocity. It should now be clear that autonomous
biochemical systems in steady state can never reach the Landauer bound on
efficiency due to the time-symmetric control penalty.

\section{Related Results}

Let's consider how the tradeoffs between accuracy and energetic-efficiency
under time-symmetric protocols compare to related approaches to thermodynamic
control, both historically and currently.

In Ref.~\cite{Bril56a}, Brillouin defines the reliability as $1/\epsilon$, where $\epsilon$ 
is the probability of reading out the wrong value due to thermal agitation.
He found that the entropy production necessary for some methods of reliable \emph{readout} is $\kB \ln(1/\epsilon)$.
It is interesting that Brillouin's readout result mirrors the same error--dissipation scaling as found in
our analysis, despite the fact that detailed fluctuation theorems, the foundations of our analysis, were still decades away.
Brillouin's readout penalty is distinct from our result, but it hints at a more general error--dissipation tradeoff  
(under particular physical constraints) that could unite the results.

Brillouin's readout penalty also appeared in disguise in Ref.~\cite{Stei77a},
which again, superficially, appears to propose a similar error--dissipation
tradeoff to what we have identified. As in Ref.~\cite{Mull76},
Ref.~\cite{Stei77a} defines a `logic operation' specifically as bit inversion,
$\zero \mapsto \one$ and $\one \mapsto \zero$---a strictly reciprocal
computation. Reference~\cite{Stei77a} thus defines the ``minimal work per logic
operation'' as the minimal work for bit inversion. Despite similarities in
appearance, it is clearly different from our result since it predicts divergent
dissipation for reliable implementation of a strictly-reciprocal computation.
In this setting, if the voltage drifts from one input to the other (a readout
error), then the wrong input will be inverted. While some computational
implementations may require further inefficiency to avoid this source of error,
it is distinct and perhaps additive to our nonreciprocity dissipation.

A discussion of reliable computing would be incomplete without mention of von
Neumann's well-known analysis of error-correction~\cite{Neum56a}. Indeed, it
might seem plausible that less reliable components can be harnessed to avoid the
error--dissipation tradeoff while nevertheless achieving the same net
reliability through error correction. However, applying the analysis above to
account for the dissipation from all components one finds that error correction
via redundancy leads to a \emph{worse} error--dissipation tradeoff. We report
on this elsewhere.

Thermodynamically inspired studies of classical reversible computing, e.g., by
Bennett, Fredkin, and Tofolli~\cite{Benn82, Fred02}, inspired much early work
on quantum computation---e.g., in Refs.~\cite{Beni82a,Zure84,Pere85}---where
logical reversibility of the unitary transformations is manifest. However,
nonreciprocity brings dissipation back to the table, even for reversible logic.
We hope it is now clear there is a necessary research program to investigate
nonreciprocity dissipation in quantum computing.

Recently, several complementary results suggest that restrictions on control
lead to additional work beyond the Landauer bound and so to dissipation. For
instance, if constrained to operate in finite time $\tau$, the dissipation
using \emph{optimal} protocols scales as the inverse $\tau^{-1}$ of the
protocol's duration. The three-way tradeoff between accuracy, speed, and energy
efficiency has been explored recently in several
settings~\cite{Muru12,Lan12,Zulk14a,Lahi16,Boyd18a}. These suggest that if
performed sufficiently slowly, an information processing protocol could achieve
the Landauer bound, with zero excess dissipation. In contrast, we found that
for time-symmetric protocols, the dissipation scales as $\ln(\epsilon^{-1})$.
Thus, one cannot achieve the Landauer bound, no matter how slowly the protocol
is executed. In this, our result is more akin to other recently explored
control restrictions, such as modularity \cite{Boyd17a, Riec18}. This too
prevents efficient extraction of nonequilibrium free energy, no matter how
slowly the protocol is executed.  

\section{Discussion: Time asymmetry?}

Another insight from the preceding is exploring time-\emph{asymmetric} control
when designing energetically efficient computation. However, is time-asymmetry
free? Often in thermodynamic control, one posits an external signal that can be
produced in any sequence, reversible or not, at no cost beyond the work-energy
imparted to the driven system. Inexpensive time asymmetry can be generated, for
example, from inertial degrees of freedom~\cite{Deff2013}.  

However, in biological systems and in many engineered nanoscale systems,
inertial degrees of freedom are not available. Noninertial batteries reliably
transforming noninertial systems will drain according to our
time-symmetric-control bounds on dissipation.

Time-symmetric control might also seem to apply, rather broadly, if one removes
the boundaries separating a driven system from its driver. Including the driver
as part of an enlarged system gives the appearance of trivially time-symmetric
driving of the composite system. Do our time-symmetric-control bounds on
dissipation then apply to any reliable computations that occur within this
enlarged system? Generically, our bounds will \emph{not} apply to such enlarged
systems because the enlarged microstate distributions $\actual[0]$ and
$\actual[\tau]$ will no longer be metastable.

This suggests that time-asymmetry and instability can both lead to energetic
advantages. Future efforts explore \emph{which} features of time-asymmetry
matter for efficiency. This helps identify those qualitative features of
time-symmetric control that must be overcome for improved efficiency.  

For time-symmetric protocols, we now ask: How does $\epsilon$ scale with $\tau$
or with the modularity of control? Generically, we do not expect a one-to-one
relationship between the error and duration. Rather, generally, there will be a
high-dimensional space of tradeoff scalings that includes speed, error,
modularity of control, robustness of information storage, energetic efficiency,
and many other computationally-relevant design restrictions.

\section{Conclusion}

We discovered that time symmetry and metastability together imply a general
error--dissipation tradeoff. The minimal work expected for a computation
$\mathcal{C}$ is the average nonreciprocity. In the low-error limit---where the
probability of error must be less than $\epsilon$---the minimum work diverges according to: 
\begin{align*}
\beta \langle W \rangle_\text{min}^\text{approx} 
= \bigl\langle \IverL \mathcal{C}(\mathcal{C}(\MSt_0)^\dagger )^\dagger
  \neq \MSt_0  \IverR \bigr\rangle_{\MSt_0} \ln(\epsilon^{-1} )
\end{align*}
Of all of this work, only the meager Landauer cost $\Delta \Shannon(\MSt_t)$,
which saturates to some finite value as $\epsilon \to 0$, can be recovered in
principle. Thus, irretrievable dissipation scales as $\ln(\epsilon^{-1} )$.
The reciprocity coefficient $ \bigl\langle \IverL
\mathcal{C}(\mathcal{C}(\MSt_0)^\dagger )^\dagger \neq \MSt_0  \IverR
\bigr\rangle_{\MSt_0}$ depends only on the deterministic computation to be
approximated. This, in turn, identifies sources of energy inefficiency in
current implementations of reliable computation. It also implies that
time-asymmetric control can allow for more efficient computation---but only
when time-asymmetric signals are a free resource.

Restricting to time-symmetric driving may seem unusual. However, with the
time-symmetric clock signal that drives modern microprocessors, it is the norm
rather than the exception in the realm of informational controls.  
Sections \ref{sec:LogicGates} and \ref{sec:NESSs} discussed the naturalness of
time-symmetry in digital computers and in genetic processing, respectively.
To get there, though, required developing several prerequisite thermodynamic results.

After developing the general thermodynamics, we analyzed the limit of
highly-reliable computing. As important examples, we then specialized to
dissipation in erasure, logic gates, and reliable biological transformations.

Overall, the results elevate time-asymmetric control to a \emph{design
principle} for efficient thermodynamic computing. This then must be added to
the growing list of recently-discovered principles, including tracking system
modularity~\cite{Boyd17a, Riec18} and logic gates whose protocols adapt to
their input~\cite{Riec18}. Further progress will turn on how these lessons are
incorporated as constraints in the principled design and search for
near-optimal finite-time protocols. An opportunity presents itself to adapt
these lessons to developing fast hyper-efficient computers in the
not-so-distant future.

\section*{Acknowledgments}
\label{sec:acknowledgments}

The authors thank Gavin Crooks, Mike DeWeese, and Chris Jarzynski for helpful
discussions and the Telluride Science Research Center for hospitality during
visits and the participants of the Information Engines Workshops there. JPC
acknowledges the kind hospitality of the Santa Fe Institute, Institute for
Advanced Study at the University of Amsterdam, and California Institute of
Technology for their hospitality during visits. This material is based upon
work supported by, or in part by, FQXi Grant number FQXi-RFP-IPW-1902,
Templeton World Charity Foundation Power of Information fellowship TWCF0337,
and U.S. Army Research Laboratory and the U.S. Army Research Office under
contracts W911NF-13-1-0390 and W911NF-18-1-0028.

\appendix
\onecolumngrid
\clearpage
\begin{center}
\large{Supplementary Materials}\\
\vspace{0.08in}
\emph{\ourTitle}\\
{\small
Paul M. Riechers, Alexander B. Boyd, Gregory W. Wimsatt, and
James P. Crutchfield
}
\end{center}

\setcounter{equation}{0}
\setcounter{figure}{0}
\setcounter{table}{0}
\setcounter{page}{1}
\makeatletter
\renewcommand{\theequation}{S\arabic{equation}}
\renewcommand{\thefigure}{S\arabic{figure}}
\renewcommand{\thetable}{S\arabic{table}}

\vspace{-0.0in}
The Supplementary Material derives a general bound on dissipated work in
thermodynamic computing, gives a transition-specific fluctuation theorem,
details the four cases of time-symmetric memories, and analyzes the minimal
dissipation in the time-symmetric universal NAND gate.

\vspace{-0.2in}
\section{Dissipated-work bound for computation}
\label{sec:CoarseGrainingDerivation_Computation}
\vspace{-0.2in}

By the information-processing inequality \cite{Cove06a}, the relative entropy
between two distributions is nonincreasing under a general channel applied to
both distributions. This has important implications for thermodynamic entropy
production.

For our main results, we consider the case where we observe memory states
$\MSet$, where $\MSet$ is a partitioning of the microstates $\SSet$, at the
initial and final times. The random variable $\MSt_t$ for the memory state at
time $t$ is completely determined by the random variable for the microstate
$\St_t$ via the memory coarse-graining function $f_{\MSt} \colon \SSet \to
\MSet$. In the following, we also invoke an even coarser graining of
microstate trajectories which takes microstate trajectories to the pair of
starting and ending memory states: $f_{\MSt_0, \MSt_\tau} (\sseq) = \bigl(
f_{\MSt} (s_0) , f_{\MSt} (s_\tau) \bigr)$.

We begin by deriving a bound on dissipation in terms of a general partitioning $\MSet$.

Note that the two probability distributions $\rho(\sseq) \equiv \Pr_{\drive}
\bigl( \St_{0:\tau} = \sseq | \St_0 \sim \actual[0] \bigr)$ and $\rho^R(\sseq)
\equiv \Pr_{\smallReverse(\drive)} \bigl( \St_{0:\tau} = \Reverse(\sseq) |
\St_0 \sim \actual[\tau]^\dagger \bigr)$ each have the same support $\sseq \in
{\SSet}^{[0,\tau]}$. From Eq.~\eqref{eq:WdissGeneralMicro}, the expected
dissipated work is: $\braket{\Wdiss} = \kB T \, \DKL{ \rho }{ \rho^R }$.

Now, consider the coarse-grained marginal distributions:
$\boldsymbol{P}(m,m') = \sum_{\sseq} \delta_{s_0 \in m} \delta_{s_\tau \in m'}
\, \rho(\sseq)$ and $\boldsymbol{Q}(m,m') = \sum_{\sseq} \delta_{s_0
\in m} \delta_{s_\tau \in m'} \, \rho^R(\sseq)$.

Since the same coarse graining kernel from $ {\SSet}^{[0,\tau]}$ to $\MSet^2$
is applied to each of the distributions $\rho$ and $\rho^R$, the
data-processing inequality guarantees that:
$\DKL{\boldsymbol{P}}{\boldsymbol{Q}} \leq \DKL{\rho}{\rho^R}$.

Moreover, note that:
\vspace{-0.1in}
\begin{align*}
\boldsymbol{P}(m,m') 
&= \sum_{\sseq} \delta_{s_0 \in m} \delta_{s_\tau \in m'} 
  \Pr_{\drive} \bigl( \St_{0:\tau} = \sseq | \St_0 \sim \actual[0] \bigr) \\
&= \Pr_{\drive} \bigl( \MSt_0 = m, \MSt_\tau = m' | \St_0 \sim \actual[0] \bigr) ~,
\vspace{-0.1in}
\end{align*}
and:
\vspace{-0.1in}
\begin{align*}
\boldsymbol{Q}(m,m') 
  & = \!\! \sum_{\sseq} \delta_{s_0 \in m} \delta_{s_\tau \in m'}  
  \!\! \Pr_{\smallReverse(\drive)} \!\! \bigl( \St_{0:\tau} = \Reverse(\sseq) | \St_0 \sim \actual[\tau]^\dagger \bigr) \\
  & = \Pr_{\smallReverse (\drive)}
  \!\! \bigl( \St_0 \in m'^{\dagger}, \St_\tau \in m^\dagger | \St_0 \sim \actual[\tau]^\dagger \bigr)
  ~.
\vspace{-0.1in}
\end{align*}

Hence, this shows in effect the dissipated work is always lower-bounded by a
function of the net transition probabilities between memory states:
\vspace{-0.1in}
\begin{align}
\beta \braket{\Wdiss} &=  \DKL{
\Pr_{\drive} \bigl( \St_{0:\tau} = \sseq | \St_0 \sim \actual[0] \bigr) 
}{
\Pr_{ \smallReverse(\drive)} \bigl( \St_{0:\tau} = \Reverse(\sseq) | \St_0 \sim \actual[\tau]^\dagger \bigr)
}
\nonumber \\
& \geq
\DKL{
\Pr_{\drive} \bigl( \MSt_0 = m, \MSt_\tau = m' | \St_0 \sim \actual[0] \bigr) 
}{
\Pr_{\smallReverse (\drive)} \bigl( \St_0 \in m'^{\dagger}, \St_\tau \in m^\dagger | \St_0 \sim \actual[\tau]^\dagger \bigr)
}~.
\label{eq:avgWdissGen_mdagninMSet}
\vspace{-0.1in}
\end{align}
Equation~\eqref{eq:avgWdissGen_mdagninMSet} is valid even when $m^\dagger \notin \MSet$.

When $m^\dagger \in \MSet$ for all $m \in \MSet$, Eq.~\eqref{eq:avgWdissGen_mdagninMSet} reduces to:
\vspace{-0.1in}
\begin{align}
\beta \langle \Wdiss \rangle \geq \DKL{\Pr_{\drive} \bigl( \MSt_0=m, \MSt_\tau=m' | \St_0 \sim \actual[0] \bigr) }{\Pr_{\smallReverse(\drive)} \bigl( \MSt_0=m'^\dagger, \MSt_\tau=m^\dagger  | \St_0 \sim \actual[\tau]^\dagger \bigr) }
\label{eq:avgWdissGeneralMacro_3}
  ~.
\vspace{-0.1in}
\end{align}
Equation~\eqref{eq:avgWdissGeneralMacro_3} will be broadly applicable to
computations performed on positional, configurational, or typical magnetic
memory systems. Indeed, there are many standard types of memory states which
satisfy $m^\dagger \in \MSet$, so that Eq.~\eqref{eq:avgWdissGeneralMacro_3} is
applicable.

For example, positionally partitioned states are often used to
store information in systems with potential energy minima at different spatial
locations. (As implemented in, for example, Maxwell demon experiments via laser
ion traps, also called `optical tweezers' when the location of these traps is
dynamically controlled.) These positional memories have a clear definition of
time reversal in which $m^\dagger=m$. This follows since if
$s=(\vec{q},\vec{\wp})$ is in $m$, then so is $s^\dagger=(\vec{q},-\vec{\wp})$.
Another form of memory state utilizes magnetic spins, for which the time
reversal flips all microstate spins---spin-up maps to spin-down upon
time-reversal, and vice versa. In the case of bistable magnetic memory
elements, the time-reversal of any memory is also a valid memory $m^\dagger \in
\MSet$, although $m^\dagger \neq m$. Thus, for both of these forms of memory,
the reversal of a memory state is an element of the partitioning, so that
Eq.~\eqref{eq:avgWdissGeneralMacro_3} is applicable.  

Further decomposing Eq.~\eqref{eq:avgWdissGeneralMacro_3}, we find:
\vspace{-0.1in}
\begin{align}
& \beta \braket{\Wdiss} 
\geq \DKL
{\Pr_{\drive} \bigl( \MSt_0 = m, \MSt_\tau = m' | \St_0 \sim \actual[0] \bigr)}
{\Pr_{\smallReverse (\drive)} \bigl( \MSt_0 = m'^\dagger, \MSt_\tau = m^\dagger | \St_0 \sim \actual[\tau]^\dagger \bigr)
}
  \nonumber \\
& = \sum_{m, m' \in \MSet}
\Pr_{\drive} \bigl( \MSt_0 = m, \MSt_\tau = m' | \St_0 \sim \actual[0] \bigr) \ln \left( \frac{\Pr_{\drive} \bigl( \MSt_0 = m, \MSt_\tau = m' | \St_0 \sim \actual[0] \bigr) }{ \Pr_{\smallReverse (\drive)} \bigl(\MSt_0 = m'^\dagger, \MSt_\tau = m^\dagger| \St_0 \sim \actual[\tau]^\dagger \bigr) } \right)
  \nonumber \\
& = \sum_{m, m' \in \MSet}
\Pr_{\drive} \bigl( \MSt_0 = m, \MSt_\tau = m' | \St_0 \sim \actual[0] \bigr)
\nonumber \\
& \qquad \qquad \qquad \times \ln \left( \frac{ \Pr_{\drive} \bigl( \MSt_0 = m | \St_0 \sim \actual[0] \bigr) \Pr_{\drive} \bigl( \MSt_\tau = m' | \St_0 \sim \actual[0] , \MSt_0 = m \bigr)  }{ \Pr_{\smallReverse (\drive)} \bigl(\MSt_0= m'^\dagger | \St_0 \sim \actual[\tau]^\dagger \bigr) \Pr_{\smallReverse (\drive)} \bigl( \MSt_\tau=m^\dagger | \St_0 \sim \actual[\tau]^\dagger , \MSt_0 = m'^\dagger \bigr) } \right) 
  \nonumber \\
  & = \sum_{m, m' \in \MSet}
  \Pr_{\drive} \bigl( \MSt_0 = m, \MSt_\tau = m' | \St_0 \sim \actual[0] \bigr) 
  \ln \left(  
\Pr_{\drive} \bigl( \MSt_0 = m | \St_0 \sim \actual[0] \bigr) \right) \nonumber \\
  & \quad - \sum_{m, m' \in \MSet}
\Pr_{\drive} \bigl( \MSt_0 = m, \MSt_\tau = m' | \St_0 \sim \actual[0] \bigr) 
  \ln \left(  
\Pr_{\smallReverse (\drive)} \bigl( \MSt_0= m'^\dagger | \St_0 \sim \actual[\tau]^\dagger \bigr)  \right) \nonumber \\
  & \quad + \sum_{m, m' \in \MSet}
\Pr_{\drive} \bigl( \MSt_0 = m, \MSt_\tau = m' | \St_0 \sim \actual[0] \bigr) 
  \ln \left( \frac{ 
\Pr_{\drive} \bigl( \MSt_\tau = m' | \St_0 \sim \actual[0] , \MSt_0 = m \bigr)  }{
\Pr_{\smallReverse (\drive)} \bigl( \MSt_\tau= m^\dagger | \St_0 \sim \actual[\tau]^\dagger , \MSt_0= m'^\dagger \bigr) } \right) 
  \nonumber \\
  & = \sum_{m \in \MSet}
\Pr_{\drive} \bigl( \MSt_0 = m | \St_0 \sim \actual[0] \bigr)
\ln \left(  
\Pr_{\drive} \bigl( \MSt_0 = m | \St_0 \sim \actual[0] \bigr) \right) \nonumber \\
& \quad - 
\sum_{m' \in \MSet}
\Pr_{\drive} \bigl( \MSt_\tau = m' | \St_0 \sim \actual[0] \bigr) \ln \left(  
\Pr_{\smallReverse (\drive)} \bigl( \MSt_0= m'^\dagger | \St_0 \sim \actual[\tau]^\dagger \bigr)  \right) \nonumber \\
& \quad + 
\sum_{m, m' \in \MSet}
\!\!\Pr_{\drive} \bigl( \MSt_0 = m | \St_0 \sim \actual[0] \bigr) 
\Pr_{\drive} \bigl( \MSt_\tau = m' | \St_0 \sim \actual[0], \MSt_0 = m \bigr)
\ln \left( \frac{ 
\Pr_{\drive} \bigl( \MSt_\tau = m' | \St_0 \sim \actual[0] , \MSt_0 = m \bigr)  }{
\Pr_{\smallReverse (\drive)} \bigl( \MSt_\tau= m^\dagger | \St_0 \sim \actual[\tau]^\dagger , \MSt_0= m'^\dagger \bigr) } \right) 
  \nonumber \\ 
&= 
\sum_{m \in \MSet}
\actual[0](m) \ln \left(  \actual[0](m) 
\right) 
\; - 
\sum_{m' \in \MSet}
\actual[\tau](m') 
\ln \left(  
{\actual[\tau]^\dagger}(m'^\dagger) 
\right) 
\nonumber \\
& \quad + 
\sum_{m, m' \in \MSet}
\actual[0](m) 
\Pr_{\drive} \bigl( \MSt_\tau = m' | \St_0 \sim \actual[0]^{(m)}  \bigr)
\ln \left( \frac{ 
\Pr_{\drive} \bigl( \MSt_\tau = m' | \St_0 \sim \actual[0]^{(m)} \bigr)  }{
\Pr_{\smallReverse (\drive)} \bigl( \MSt_\tau= m^\dagger | \St_0 \sim \actual[\tau]^{\dagger (m'^\dagger)} \bigr) } \right) 
  \nonumber \\ 
&= 
- \Shannon (\MSt_0) + \Shannon (\MSt_\tau)
+ 
\sum_{m, m' \in \MSet}
\actual[0](m) d(m,m') 
  \nonumber \\
& = \Delta \Shannon (\MSt_t)
+ 
\sum_{m, m' \in \MSet}
\actual[0](m) d(m,m')
  ~.
\label{eq:GenericMemoryDissipationBound}
\vspace{-0.1in}
\end{align}
In the third-to-last equation above we used the fact that
${\actual[\tau]^\dagger}(m^\dagger) = {\actual[\tau]}(m)$ and we defined:
\vspace{-0.1in}
\begin{align*}
d(m,m') & \equiv 
\Pr_{\drive} \bigl(  \MSt_\tau = m' \big| \St_0 \sim \actual[0]^{(m)} \bigr)
\ln \left( \frac{\Pr_{\drive} \bigl(  \MSt_\tau = m'  \big| \St_0 \sim \actual[0]^{(m)}  \bigr) }{\Pr_{ \smallReverse (\drive)} \bigl( \MSt_\tau = m^\dagger \big| \St_0 \sim \actual[\tau]^{ \dagger (m'^\dagger)}  \bigr) } \right) ~.
\vspace{-0.1in}
\end{align*}

A more general derivation (\emph{not} assuming that $m^\dagger \in \MSet$),
fully leveraging Eq.~\eqref{eq:avgWdissGen_mdagninMSet},
yields Eq.~\eqref{eq:GenericMemoryDissipationBound} with:
\vspace{-0.1in}
\begin{align*}
d(m,m') \equiv 
\Pr_{\drive} \bigl(  \MSt_\tau = m' \big| \St_0 \sim \actual[0]^{(m)} \bigr)
  \ln \left( \frac{\Pr_{\drive} \bigl(  \MSt_\tau = m'  \big| \St_0 \sim \actual[0]^{(m)}  \bigr) }{\Pr_{ \smallReverse (\drive)} \bigl( \St_\tau \in m^\dagger \big| \St_0 \sim \actual[\tau]^{ \dagger (m'^\dagger)}  \bigr) } \right)
  ~.
\vspace{-0.1in}
\end{align*}
This generalization is useful, for example, when analyzing tristable magnetic
memory elements.

In the $m=m^\dagger$ case, where memories are stored via time-symmetric
variables, these expressions both simplify to:
\vspace{-0.1in}
\begin{align*}
d(m,m') = 
\Pr_{\drive} \bigl(  \MSt_\tau = m' \big| \St_0 \sim \actual[0]^{(m)} \bigr)
\ln \left( \frac{\Pr_{\drive} \bigl(  \MSt_\tau = m'  \big| \St_0 \sim \actual[0]^{(m)}  \bigr) }{\Pr_{ \smallReverse (\drive)} \bigl( \MSt_\tau = m \big| \St_0 \sim \actual[\tau]^{ \dagger (m')}  \bigr) } \right) 
  ~.
\vspace{-0.1in}
\end{align*}
This compares (i) the probability of transitioning from memory state $m$ to
$m'$ to (ii) the probability of returning back to $m$ upon reversal of time-odd
variables (like momentum) and subsequent reversal of the control protocol.
Above, $\actual[t]^{(m)} \equiv \delta_{\St_t \in m} \actual[t] / \sum_{s' \in
m} \actual[t](s')$ is the renormalized restriction of the microstate
distribution $\actual[t]$ to the set of microstates represented by the memory
$m$. Note that (despite the fact that ${\actual[\tau]^\dagger}(m^\dagger) =
{\actual[\tau]}(m) $), generically $\actual[\tau]^{ \dagger (m^\dagger)} \neq
\actual[\tau]^{  (m)}$. However, conveniently, metastable memory systems have: $\actual[\tau]^{ \dagger (m)} \approx {\boldsymbol{\pi}}_{x_\tau}^{ \dagger (m)}
= \LocalEq[x_\tau^\dagger]{m}$.

\vspace{-0.1in}
\section{Transition-specific fluctuation theorem}
\label{sec:TransSpecFT}
\vspace{-0.1in}

Here, we verify the transition-specific version of Eq.~\eqref{eq:WorkRequired}.
We start with the detailed fluctuation theorem for heat:
\vspace{-0.1in}
\begin{align*}
e^{\beta Q{(\sseq, \drive)}} 
= \frac{\Pr_{\drive} \bigl( \St_{0:\tau} = \sseq | \St_0 = s_0 \bigr) }{\Pr_{ \smallReverse(\drive)} \bigl( \St_{0:\tau} = \Reverse(\sseq) | \St_0 = s_\tau^\dagger \bigr)} 
  ~.
\vspace{-0.1in}
\end{align*}
We then apply the First Law of thermodynamics (energy
conservation)---$E_{x_\tau}(s_\tau) - E_{x_0}(s_0) = W - Q $---together with a
useful identity: $E_x(s) = -\kB T \ln \bigl(  \LocalEq[x]{m}(s) \bigr) +
F_x^{(m)}$, if $s \in m$. This allows us to write:
\begin{align*}
W - \Delta  F_x^{(m)} =  Q - \kB T \Delta \ln  \bigl(  \LocalEq[x]{m}(s) \bigr) ~.
\end{align*}
Then we calculate:
\begin{align*}
& \braket{ e^{- \beta (W - \Delta F_x^{(\MSt)} )}  }_{\Pr_{\drive}(\St_{0:\tau} | \MSt_\tau = m', \St_0 \sim  \LocalEq[x_0]{m})} \\
& \qquad = 
\bigl\langle e^{- \beta \bigl( Q - \kB T \Delta  \ln  \bigl(  \LocalEq[x]{\MSt}(\St) \bigr) \bigr) } \bigr\rangle_{\Pr_{\drive}(\St_{0:\tau} | \MSt_\tau = m', \St_0 \sim  \LocalEq[x_0]{m})} \\
& \qquad = 
\sum_{ s_{0:\tau} \in \SSet^{[0:\tau]} }
 \Pr_{\drive}(\St_{0:\tau} = s_{0:\tau}  |  \MSt_\tau = m', \St_0 \sim  \LocalEq[x_0]{m})
\frac{ \LocalEq[x_\tau]{m'}(s_\tau) }{ \LocalEq[x_0]{m}(s_0)  }
\frac{ \Pr_{ \smallReverse(\drive)} \bigl( \St_{0:\tau} = \Reverse(\sseq) | \St_0 = s_\tau^\dagger \bigr)  }{ \Pr_{\drive} \bigl( \St_{0:\tau} = \sseq | \St_0 = s_0 \bigr) }
 \\
& \qquad = 
\sum_{ s_{0:\tau} \in m \SSet^{(0:\tau)} m' }
 \Pr_{\drive}(\St_{0:\tau} = s_{0:\tau}  | \MSt_\tau = m', \St_0 \sim  \LocalEq[x_0]{m})
\frac{ \LocalEq[x_\tau]{m'}(s_\tau) }{ \LocalEq[x_0]{m}(s_0)  }
\frac{ \Pr_{ \smallReverse(\drive)} \bigl( \St_{0:\tau} = \Reverse(\sseq) | \St_0 = s_\tau^\dagger \bigr)  }{ \Pr_{\drive} \bigl( \St_{0:\tau} = \sseq | \St_0 = s_0 \bigr) }
 \\
 & \qquad = 
 \tfrac{1}{ \Pr_{\drive}( \MSt_\tau = m' | \St_0 \sim  \LocalEq[x_0]{m})}
\sum_{ s_{0:\tau} \in m \SSet^{(0:\tau)} m' } 
 \Pr_{\drive}(\St_{0:\tau} = s_{0:\tau} |  \St_0 \sim  \LocalEq[x_0]{m})
\frac{ \LocalEq[x_\tau]{m'}(s_\tau) }{ \LocalEq[x_0]{m}(s_0)  }
\frac{ \Pr_{ \smallReverse(\drive)} \bigl( \St_{0:\tau} = \Reverse(\sseq) | \St_0 = s_\tau^\dagger \bigr)  }{ \Pr_{\drive} \bigl( \St_{0:\tau} = \sseq | \St_0 = s_0 \bigr) }
 \\
  & \qquad = 
 \frac{1}{ \Pr_{\drive}( \MSt_\tau = m' | \St_0 \sim  \LocalEq[x_0]{m})}
\sum_{ s_{0:\tau} \in m \SSet^{(0:\tau)} m' }
\LocalEq[x_\tau]{m'}(s_\tau)
\Pr_{ \smallReverse(\drive)} \bigl( \St_{0:\tau} = \Reverse(\sseq) | \St_0 = s_\tau^\dagger \bigr) 
 \\
   & \qquad = 
 \frac{1}{ \Pr_{\drive}( \MSt_\tau = m' | \St_0 \sim  \LocalEq[x_0]{m})}
\sum_{ \smallReverse(s_{0:\tau}) \in m'^\dagger \SSet^{(0:\tau)} m^\dagger }
\LocalEq[x_\tau^\dagger]{m'^\dagger}(s_\tau^\dagger)
\Pr_{ \smallReverse(\drive)} \bigl( \St_{0:\tau} = \Reverse(\sseq) | \St_0 = s_\tau^\dagger \bigr) 
 \\
 & \qquad = 
 \frac{1}{ \Pr_{\drive}( \MSt_\tau = m' | \St_0 \sim  \LocalEq[x_0]{m})}
\sum_{ \smallReverse(s_{0:\tau}) \in m'^\dagger \SSet^{(0:\tau)} m^\dagger }
\Pr_{ \smallReverse(\drive)} \bigl( \St_{0:\tau} = \Reverse(\sseq) | \St_0 \sim \LocalEq[x_\tau^\dagger]{m'^\dagger} \bigr) 
 \\
  & \qquad = 
 \frac{1}{ \Pr_{\drive}( \MSt_\tau = m' | \St_0 \sim  \LocalEq[x_0]{m})}
\sum_{ s \in  m^\dagger }
\Pr_{ \smallReverse(\drive)} \bigl( \St_{\tau} = s | \St_0 \sim \LocalEq[x_\tau^\dagger]{m'^\dagger} \bigr) 
 \\
 & \qquad = 
 \frac{ \Pr_{ \smallReverse(\drive)} \bigl( \MSt_{\tau} = m^\dagger | \St_0 \sim \LocalEq[x_\tau^\dagger]{m'^\dagger} \bigr)  }{\Pr_{\drive}( \MSt_\tau = m' | \St_0 \sim  \LocalEq[x_0]{m}) }
  ~,
\end{align*}
where we used the fact that $\LocalEq[x_\tau]{m'}(s_\tau) = \LocalEq[x_\tau^\dagger]{m'^\dagger}(s_\tau^\dagger)$.

Altogether, this leads to a useful transition-specific fluctuation theorem:
\vspace{-0.1in}
\begin{align*}
\braket{e^{-\beta W}}_{\Pr_{\drive}(\St_{0:\tau} | \MSt_\tau = m', \St_0 \sim  \LocalEq[x_0]{m})}
= e^{-\beta \Delta F_x^{(\MSt)}} \frac{
 \Pr_{\smallReverse(\drive)} \bigl(  \MSt_\tau = m^\dagger  \big| \St_0 \sim \boldsymbol{\pi}_{x_\tau^\dagger}^{(m'^\dagger)}   \bigr) } { \Pr_{\drive} \bigl(  \MSt_\tau = m'  \big| \St_0 \sim \LocalEq[x_0]{m}  \bigr)}~.
\vspace{-0.1in}
\end{align*}
By Jensen's inequality, this yields:
\vspace{-0.1in}
\begin{align*}
\braket{W}_{\Pr_{\drive}(\St_{0:\tau} | \MSt_\tau = m', \St_0 \sim  \LocalEq[x_0]{m})} 
\geq   \Delta F_x^{(\MSt)}  
+ \kB T \ln 
 \frac{\Pr_{\drive}( \MSt_\tau = m' | \St_0 \sim  \LocalEq[x_0]{m}) }{ \Pr_{ \smallReverse(\drive)} \bigl( \MSt_{\tau} = m^\dagger | \St_0 \sim \LocalEq[x_\tau^\dagger]{m'^\dagger} \bigr)  } ~.
\vspace{-0.1in}
\end{align*}
Assuming time-symmetric control, this then implies that the minimal average
work required of a memory transition is:
\vspace{-0.1in}
\begin{align*}
W_\text{min}^{t\text{-sym}}(m \to m') = F_{x_0}^{(m')} - F_{x_0}^{(m)} + \kB T \NR (m \to m') ~.
\vspace{-0.1in}
\end{align*}

These results exhibit similarity with those of Ref.~\cite{Engl13}, although the
appearance of the local-equilibrium free energies is a new feature here. This
has important implications for the work required of metastable memory
transitions.

\section{Four cases}
\label{sec:4cases}

To evaluate Eq.~\eqref{eq:dmmSimplified} there are four cases to consider.
Assuming time-symmetric memories (such that $m^\dagger=m$), the four cases
depend on whether $\mathcal{C}(m) = m'$ or $\neq m'$, and on whether
$\mathcal{C}(m') = m$ or $\neq m$. Whatever implementation is used for the
computation, it will result in some \emph{actual} probability of error for each
of the intended transitions: $\Pr_{\drive} \bigl(  \MSt_\tau = \mathcal{C}(m)
\big| \MSt_0 = m \bigr) = 1 - \epsilon_{m}$. The reliability
design constraint is that $\epsilon_{m} \leq \epsilon$ for all possible initial
memories $m$, and that $\epsilon \ll 1$, so that errors are very improbable.

Given some implementation $\drive$, the actual probability of making a
particular \emph{accidental} memory transition is $\epsilon_{m \rightarrow m'}$ for $m'
\neq \mathcal{C}(m)$. Since $\sum_{m' \in \MSet \setminus \{ \mathcal{C}(m) \}
} \epsilon_{m \rightarrow m'} = \epsilon_{m} \leq \epsilon$, we must have that $0 \leq
\epsilon_{m\rightarrow m'} \leq \epsilon_{m} \leq \epsilon$. Let us use this to evaluate the
four possible cases for $d(m,m')$:
\begin{enumerate}
      \setlength{\topsep}{-5pt}
      \setlength{\itemsep}{-5pt}
      \setlength{\parsep}{-5pt}
      \setlength{\labelwidth}{5pt}
\item $\mathcal{C}(m) = m'$; $\mathcal{C}(m') = m$:
\begin{align*}
d^{(1)}(m,m') 
&= (1 - \epsilon_{m}) \ln \left( \frac{1-\epsilon_{m}}{1- \epsilon_{m'}} \right) \\
&\geq (1 - \epsilon_{m}) \ln \left( 1-\epsilon_{m}  \right) \\
&\geq (1 - \epsilon) \ln \left( 1- \epsilon  \right) \\
&= (1 - \epsilon)  \left( - \epsilon - \tfrac{1}{2} \epsilon^2 - \tfrac{1}{3} \epsilon^3 - \dots \right) \\
&= - \epsilon + \tfrac{1}{2} \epsilon^2 + \tfrac{1}{6} \epsilon^3 + \dots \\
& \geq - \epsilon + \tfrac{1}{2} \epsilon^2  \\
& \geq - \epsilon \\
& \approx 0
\end{align*}
Similarly,
\begin{align*}
d^{(1)}(m,m') 
&= (1 - \epsilon_{m}) \ln \left( \frac{1-\epsilon_{m}}{1- \epsilon_{m'}} \right) \\
&\leq - \ln(1-\epsilon_{m'} ) \\
&\leq - \ln(1-\epsilon ) \\
&= \epsilon + \tfrac{1}{2} \epsilon^2 + \tfrac{1}{3} \epsilon^3 + \dots \\
& \approx 0
\end{align*}
So $- \epsilon \leq d^{(1)}(m,m') \leq \epsilon + \tfrac{1}{2} \epsilon^2 + \mathcal{O}(\epsilon^3)$.
\item $\mathcal{C}(m) = m'$; $\mathcal{C}(m') \neq m$:
\begin{align*}
d^{(2)}(m,m') 
&= (1 - \epsilon_{m}) \ln \left( \frac{1-\epsilon_{m}}{\epsilon_{m' \rightarrow m}} \right) \\
&\geq  (1-\epsilon) \ln \left( \frac{1-\epsilon}{\epsilon_{m' \rightarrow m}} \right)  \\
& \geq (1-\epsilon) \ln \left( \frac{1-\epsilon}{\epsilon} \right)  \\
& \approx \ln \left( 1 / \epsilon \right)
\end{align*}
We also have:
\begin{align*}
d^{(2)}(m,m') 
&= (1 - \epsilon_{m}) \ln \left( \frac{1-\epsilon_{m}}{\epsilon_{m'\rightarrow m}} \right) \\
&\leq \ln (1/ \epsilon_{m' \rightarrow m}) \\
\end{align*}
So, $\ln(\epsilon^{-1}) \lesssim d^{(2)}(m,m') \leq  \ln (\epsilon_{m' \rightarrow m}^{-1})  $.
\item $\mathcal{C}(m) \neq m'$; $\mathcal{C}(m') = m$:
\begin{align*}
d^{(3)}(m,m') 
&= \epsilon_{m \rightarrow m'} \ln \left( \frac{\epsilon_{m \rightarrow m'}}{1- \epsilon_{m'}} \right) \\
& > \epsilon_{m \rightarrow m'} \ln \epsilon_{m \rightarrow m'} \\
& \approx 0
\end{align*}
Also:
\begin{align*}
d^{(3)}(m,m') 
&= \epsilon_{m,m'} \ln \left( \frac{\epsilon_{m,m'}}{1- \epsilon_{m'}} \right) \\
& < 0 ~.
\end{align*}
So, $ \epsilon_{m \rightarrow m'} \ln \epsilon_{m \rightarrow m'} < d^{(3)}(m,m') < 0$.
\item $\mathcal{C}(m) \neq m'$; $\mathcal{C}(m') \neq m$:
\begin{align*}
d^{(4)}(m,m') 
&= \epsilon_{m \rightarrow m'} \ln \left( \frac{\epsilon_{m \rightarrow m'}}{\epsilon_{m' \rightarrow m}} \right)
\\ & < \epsilon_{m \rightarrow m'} \ln \left( \frac{1}{\epsilon_{m' \rightarrow m}} \right)
\\ & \approx 0 ~,
\end{align*}
assuming that the errors $\epsilon_{m \rightarrow m'}$ and $\epsilon_{m' \rightarrow m}$ are on the same order of magnitude.  Also:
\begin{align*}
d^{(4)}(m,m') 
&= \epsilon_{m \rightarrow m'} \ln \left( \frac{\epsilon_{m \rightarrow m'}}{\epsilon_{m' \rightarrow m}} \right) \\
  & > \epsilon_{m \rightarrow m'} \ln \epsilon_{m \rightarrow m'} \\
  & \approx 0
  ~.
\end{align*}
So, $ \epsilon_{m \rightarrow m'} \ln \epsilon_{m \rightarrow m'} < d^{(4)}(m,m') <- \epsilon_{m \rightarrow m'} \ln \epsilon_{m' \rightarrow m}$.
\end{enumerate}

Case 2 is the only one resulting in divergent dissipation with increasing reliability.

However, due to the ratio of error rates in Case 4, its behavior in the
low-error limit deserves further analysis.

On the one hand, suppose $\epsilon_{m' \rightarrow m}
> \epsilon_{m \rightarrow m'}$. Let $r \equiv \epsilon_{m \rightarrow
m'} / \epsilon_{m' \rightarrow m} < 1$. Then:
\begin{align*}
\vspace{-0.1in}
d^{(4)}(m, m')
= \epsilon_{m \rightarrow m'} \ln \frac{\epsilon_{m \rightarrow m'}}{\epsilon_{m' \rightarrow m}}
= \epsilon_{m' \rightarrow m} r \ln r
 ~.
\vspace{-0.1in}
\end{align*}
If we define $f(r) = r \ln r$, then $f'(r) = \ln r + 1$.
And so, $f'(r^*) = 0$ gives $r^* = e^{-1}$ and $f(r^*) = - e^{-1}$.
Since $f(0) = f(1) = 0$, $f$'s local minimum is at $e^{-1}$. So:
\vspace{-0.1in}
\begin{align*}
d^{(4)}(m, m') = \epsilon_{m' \rightarrow m} r \ln r \geq -\epsilon_{m' \rightarrow m} e^{-1}
 ~.
\vspace{-0.1in}
\end{align*}
However, $\epsilon \geq \epsilon_{m' \rightarrow m}$, so
$d^{(4)}(m, m') \geq - \epsilon e^{-1}$.
And, since $f(r) \leq 0$ for $0 \leq r \leq 1$,
we have $0 \geq d^{(4)}(m, m') \geq -\epsilon e^{-1}$.

If, on the other hand, $\epsilon_{m \to m'} > \epsilon_{m' \to m}$, then Case 4
will imply some extra dissipation, although it will still be on the order of
$\epsilon$ so long as $\epsilon_{m \to m'} $ and $\epsilon_{m' \to m}$ are
within several orders of magnitude.

\vspace{-0.1in}
\section{Minimal dissipation in the time-symmetric universal NAND gate}
\label{sec:NAND}
\vspace{-0.1in}

Let us consider the implications for the common logic gates that serve as the
building blocks for practical computers. Recall that the simple NAND gate is
sufficient for universal computation. It is therefore worthwhile to consider
what dissipation is commonly incurred in such important logic gates---and
consider how this dissipation can be avoided.

To address this, consider a physical implementation of the memory components of
a NAND gate, where we explicitly consider two memory elements whose memory
states are to be used as the input for the NAND gate and another memory element
that will store the value of the output. We assume that only the output is
over-written during the computation---the input memory states may be kept
around for later use.

Note that this is already a particular physical model of the NAND
computation---indeed, alternatives exist such as storing the output in the
location of one of the former inputs by over-writing one of the inputs.
However, we analyze the proposed two-input--one-output model here since it is
arguably the most relevant to the typical desired use of a NAND gate. Other
ancillary memory elements may be used in the computation as in
Ref.~\cite{Owen17} but, since they will be returned to their original state by
the computation's end, these ancillary memories do not need to result in any
explicit additional dissipation dependence---although they can influence the
dissipation through their implicit effect on the net errors in operating on the
memory-elements of interest---and can be left implicit in the self-consistency
of the current analysis.

Each of the three explicitly-considered memory elements is assumed to be
bistable; i.e., each has two metastable regions of state space.
Let the
microstate of each memory element be specified by its position in the interval
$(-\pi, \pi]$. Between computations, including at $t=0$ and $t=\tau$, the
metastable regions for each memory element are $\zero \equiv (-\pi, 0]$ and
$\one \equiv (0, \pi]$ which gives a natural partition for the memory states.

The microstate of the memory system at any time $t$ can be treated as a
composite random variable $\St_t = (\St_t^{(\text{in}_1)},
\St_t^{(\text{in}_2)}, \St_t^{(\text{out})})$ with $\St_t^{(\cdot)} \in (-\pi,
\pi]$. Similarly, the memory state is the composite random variable $\MSt_t =
(\MSt_t^{(\text{in}_1)}, \MSt_t^{(\text{in}_2)}, \MSt_t^{(\text{out})})$ with
$\MSt_t^{(\cdot)} \in \{ \zero, \one \}$ corresponding to the two metastable
regions for each memory element. Thus, the joint state-space $\SSet =
\mathbb{R}_{(-\pi, \pi]}^3$ has eight metastable regions, which we identify as
the joint memory system's eight memory states: $\MSet = \{ m_{\zero \zero
\zero}, m_{\zero \zero \one}, m_{\zero \one \zero}, \dots m_{\one \one \one}
\}$ where each memory state is labeled according to its corresponding region of
state-space: $m_{j k \ell} = \bigl\{ s \in \SSet : \, s^{(\text{in}_1)}
\in (-\pi + j \pi, j \pi], \, s^{(\text{in}_2)} \in (-\pi + k \pi, k \pi], \,
s^{(\text{out})} \in (-\pi + \ell \pi, \ell \pi]  \bigr\}$. That is, each of
the memory states is one of the octants of state space.

\begin{figure}[t]
\includegraphics[width = 0.75\linewidth]{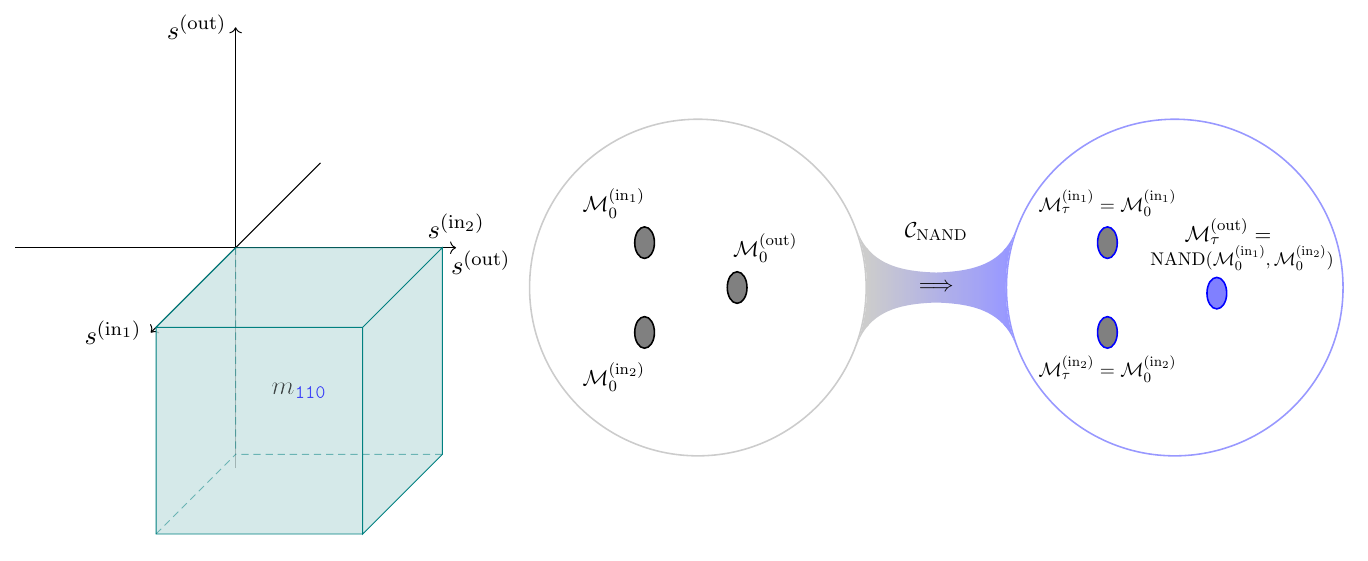}
\caption{Composite state space, memory states, and computation associated with
	the logical NAND operation.
	[Reproduced with permission (with minor modification) from Ref.~\cite{Riec18}.]
	}
\end{figure}

$\mathcal{C}_\text{NAND}$ is given by the mappings:
\begin{align*}
\zero \zero \zero & \mapsto \zero \zero \one ~~~~
\zero \zero \one \mapsto \zero \zero \one ~~~~ 
\zero \one \zero \mapsto \zero \one \one ~~~~
\zero \one \one \mapsto \zero \one \one ~~~~
\one \zero \zero \mapsto \one \zero \one ~~~~
\one \zero \one \mapsto \one \zero \one ~~~~
\one \one \zero \mapsto \one \one \zero ~~~~
\one \one \one \mapsto \one \one \zero \\
\end{align*}

For arbitrary initial correlation among the memory elements before the
computation, the change in memory entropy is:
\begin{align*}
\Delta & \Shannon (\MSt_t)  \\
= &
\Shannon \bigl( \MSt_\tau^{(\text{in}_1)} , \MSt_\tau^{(\text{in}_2)} , \MSt_\tau^{(\text{out})} \bigr) 
\!\! -\!\!  
\Shannon \bigl( \MSt_0^{(\text{in}_1)} , \MSt_0^{(\text{in}_2)} , \MSt_0^{(\text{out})} \bigr) \\
\approx &
\Shannon \bigl( \MSt_0^{(\text{in}_1)} , \MSt_0^{(\text{in}_2)} , \MSt_\tau^{(\text{out})} \bigr) 
\!\! -\!\!  
\Shannon \bigl( \MSt_0^{(\text{in}_1)} , \MSt_0^{(\text{in}_2)} , \MSt_0^{(\text{out})} \bigr) \\
\approx &
\Shannon \bigl( \MSt_0^{(\text{in}_1)} , \MSt_0^{(\text{in}_2)} \bigr) 
\!\! -\!\!  
\Shannon \bigl( \MSt_0^{(\text{in}_1)} , \MSt_0^{(\text{in}_2)} , \MSt_0^{(\text{out})} \bigr) \\
= & - 
\Shannon \bigl( \MSt_0^{(\text{out})} \big| \MSt_0^{(\text{in}_1)} , \MSt_0^{(\text{in}_2)} \bigr)
  ~,
\end{align*}
where we made the approximation that $\epsilon$ is very small. The entropy
$\Shannon \bigl( \MSt_0^{(\text{out})} \big| \MSt_0^{(\text{in}_1)} ,
\MSt_0^{(\text{in}_2)} \bigr) $ is the Landauer bound for the NAND gate---a
result of interest in its own right. Here, we use it to provide the $\Delta
\Shannon (\MSt_t) $ needed for our calculation of dissipation for
time-symmetric implementations of NAND.

It should be noted that, by the definition of $\mathcal{C}_\text{NAND}$, half
of the memory states map to themselves. The other half never get mapped back to
themselves upon iteration of the computation. The reciprocity coefficient is
therefore:
\begin{align*}
\bigl\langle \IverL
& \mathcal{C}_\text{NAND} (\mathcal{C}_\text{NAND}(\MSt_0) ) \neq \MSt_0 \IverR
\bigr\rangle_{\MSt_0} \\
& \quad = \actual(\zero \zero \zero) + \actual(\zero
\one \zero) + \actual(\one \zero \zero) + \actual(\one \one \one)
  ~.
\end{align*}
Together with Eq.~\eqref{eq:MainScalingResult}, these give the general lower
bound for NAND gate dissipation when a time-symmetric protocol is used and the
memory elements are initially correlated.

In the simpler case where the memory elements are initiated with individual
biases towards $\zero$ or $\one$, but are otherwise uncorrelated with each
other, the NAND dissipation somewhat simplifies, since then:
\begin{align*}
\Delta \Shannon
(\MSt_t) & \approx
  - \Shannon \bigl( \MSt_0^{(\text{out})} \big|
\MSt_0^{(\text{in}_1)} , \MSt_0^{(\text{in}_2)} \bigr) \\
  & = - \Shannon \bigl( \MSt_0^{(\text{out})} \bigr) \\
  & = - \boldsymbol{H}_\text{b} ( b_0^{(\text{out})} )
  ~,
\end{align*}
where $b_0^{(\text{out})}$ is the initial bias of the output bit to be
overwritten.

In this latter uncorrelated case, the time-symmetric NAND dissipation resembles
the dissipation that would be expected simply from bit erasure of the output
bit. The primary difference is that the probability of overwriting the output
bit now depends intricately on a function $\bigl\langle \IverL
\mathcal{C}_\text{NAND}(\mathcal{C}_\text{NAND}(\MSt_0) ) \neq \MSt_0 \IverR
\bigr\rangle_{\MSt_0}$ of the distribution over the joint memory state of all
memory elements which is not separable even when the joint distribution over
memory elements is separable.

It should be noted that, even in the case of biased but uncorrelated initialization of the memory, the initialization statistics would need to be taken into account to make the time-symmetric physical implementation as thermodynamically efficient as possible. Interestingly, even in the case of time-asymmetric protocols, it has been found that the initialization statistics must be taken into account for thermodynamic efficiency of the NAND gate~\cite{Riec18}. Thus, the impetus to match the input statistics is somewhat of a separate generic issue in addition to the lesson of extra dissipation that generically accompanies time-symmetric implementation of reliable computation.


\begin{thebibliography}{10}

\bibitem{Benn82}
C.~H. Bennett.
\newblock Thermodynamics of computation---{A} review.
\newblock {\em Intl. J. Theo. Phys.}, 21:905, 1982.

\bibitem{Maro09}
O.~J.~E. Maroney.
\newblock Generalizing {Landauer's} principle.
\newblock {\em Phys. Rev. E}, 79:031105, Mar 2009.

\bibitem{Saga14}
T.~Sagawa.
\newblock Thermodynamic and logical reversibilities revisited.
\newblock {\em J. Stat. Mech. Th. Exp.}, 2014(3):P03025, 2014.

\bibitem{Parr15a}
J.~M.~R. Parrondo, J.~M. Horowitz, and T.~Sagawa.
\newblock Thermodynamics of information.
\newblock {\em Nature Physics}, 11(2):131--139, February 2015.

\bibitem{Riec18}
P.~M. Riechers.
\newblock Transforming metastable memories: The nonequilibrium thermodynamics
  of computation.
\newblock In D.~Wolpert, C.~Kempes, P.~Stadler, and J.~Grochow, editors, {\em
  The Energetics of Computing in Life and Machines}. SFI Press, 2019.

\bibitem{Sala83}
P.~Salamon and R.~S. Berry.
\newblock Thermodynamic length and dissipated availability.
\newblock {\em Phys. Rev. Lett.}, 51:1127--1130, Sep 1983.

\bibitem{Siva12}
D.~A. Sivak and G.~E. Crooks.
\newblock Thermodynamic metrics and optimal paths.
\newblock {\em Phys. Rev. Lett.}, 108:190602, May 2012.

\bibitem{Zulk14}
P.~R. Zulkowski and M.~R. DeWeese.
\newblock Optimal finite-time erasure of a classical bit.
\newblock {\em Phys. Rev. E}, 89(5):052140, 2014.

\bibitem{Bona14}
M.~VS. Bonan{\c{c}}a and S.~Deffner.
\newblock Optimal driving of isothermal processes close to equilibrium.
\newblock {\em J. Chem. Phys.}, 140(24):244119, 2014.

\bibitem{Mand16}
D.~Mandal and C.~Jarzynski.
\newblock Analysis of slow transitions between nonequilibrium steady states.
\newblock {\em J. Stat. Mech. Th. Exp.}, 2016(6):063204, 2016.

\bibitem{Boyd18a}
A.~B. Boyd, A.~Patra, C.~Jarzynski, and J.~P. Crutchfield.
\newblock Shortcuts to thermodynamic computing: The cost of fast and faithful
  information processing.
\newblock arXiv:1812.11241.

\bibitem{Chan92}
A.~P. {Chandrakasan}, S.~{Sheng}, and R.~W. {Brodersen}.
\newblock Low-power {CMOS} digital design.
\newblock {\em IEEE J. Solid-State Circ.}, 27(4):473--484, April 1992.

\bibitem{Iyer02}
A.~Iyer and D.~Marculescu.
\newblock Power efficiency of voltage scaling in multiple clock, multiple
  voltage cores.
\newblock In {\em Proc. 2002 IEEE/ACM Intl. Cong. Computer-aided Design}, pages
  379--386. ACM, 2002.

\bibitem{Seld10}
J.~S. Seldenthuis, F.~Prins, J.~M. Thijssen, and H.~S.~J. van~der Zant.
\newblock An all-electric single-molecule motor.
\newblock {\em ACS nano}, 4(11):6681--6686, 2010.

\bibitem{Astu17}
R.~D. Astumian.
\newblock How molecular motors work--insights from the molecular machinist's
  toolbox: {The Nobel Prize in Chemistry} 2016.
\newblock {\em Chemical Sci.}, 8(2):840--845, 2017.

\bibitem{Brow17}
A.~I. Brown and D.~A. Sivak.
\newblock Toward the design principles of molecular machines.
\newblock {\em arXiv:1701.04868}.

\bibitem{Zhan18}
L.~Zhang, V.~Marcos, and D.~A. Leigh.
\newblock Molecular machines with bio-inspired mechanisms.
\newblock {\em Proc. Natl. Acad. Sci.}, 115(38):9397--9404, 2018.

\bibitem{Feng08}
E.~H. Feng and G.~E. Crooks.
\newblock Length of time's arrow.
\newblock {\em Phys. Rev. Lett.}, 101:090602, Aug 2008.

\bibitem{Land61a}
R.~Landauer.
\newblock Irreversibility and heat generation in the computing process.
\newblock {\em IBM J. Res. Develop.}, 5(3):183--191, 1961.

\bibitem{Croo99}
G.~E. Crooks.
\newblock Entropy production fluctuation theorem and the nonequilibrium work
  relation for free energy differences.
\newblock {\em Phys. Rev. E}, 60:2721--2726, 1999.

\bibitem{Jarz00}
C.~Jarzynski.
\newblock Hamiltonian derivation of a detailed fluctuation theorem.
\newblock {\em J. Stat. Physics}, 98(1-2):77--102, 2000.

\bibitem{Seif05}
U.~Seifert.
\newblock Entropy production along a stochastic trajectory and an integral
  fluctuation theorem.
\newblock {\em Phys. Rev. Lett.}, 95:040602, Jul 2005.

\bibitem{Espo11a}
M.~Esposito and C.~Van den Broeck.
\newblock Second law and {Landauer} principle far from equilibrium.
\newblock {\em Eur. Phys. Lett.}, 95(4), 2011.

\bibitem{Cove06a}
T.~M. Cover and J.~A. Thomas.
\newblock {\em Elements of Information Theory}.
\newblock Wiley-Interscience, New York, second edition, 2006.

\bibitem{Espo12}
M.~Esposito.
\newblock Stochastic thermodynamics under coarse graining.
\newblock {\em Phys. Rev. E}, 85(4):041125, 2012.

\bibitem{Jarz06a}
C.~Jarzynski.
\newblock Rare events and the convergence of exponentially averaged work
  values.
\newblock {\em Phys. Rev. E}, 73:046105, Apr 2006.

\bibitem{Kawa07}
R.~Kawai, J.~M.~R. Parrondo, and C.~Van den Broeck.
\newblock Dissipation: The phase-space perspective.
\newblock {\em Phys. Rev. Lett.}, 98:080602, 2007.

\bibitem{Gome08a}
A.~Gomez-Marin, J.~M.~R. Parrondo, and C.~{Van den Broeck}.
\newblock Lower bounds on dissipation upon coarse graining.
\newblock {\em Phys. Rev. E}, 78(1):011107, 2008.

\bibitem{Rold10b}
E.~Roldan and J.~M.~R. Parrondo.
\newblock Estimating dissipation from single stationary trajectories.
\newblock {\em Phys. Rev. Lett.}, 105(150607), 2010.

\bibitem{Horo09a}
J.~Horowitz and C.~Jarzynski.
\newblock Illustrative example of the relationship between dissipation and
  relative entropy.
\newblock {\em Phys. Rev. E}, 79(021106), 2009.

\bibitem{Rold10a}
E.~Roldan and J.~M.~R. Parrondo.
\newblock {\em Estimating dissipation from single stationary trajectories}.
\newblock PhD thesis, Universidad Complutense de Madrid, 28040 Madrid, Spain,
  September 2010.

\bibitem{Jun14a}
Y.~Jun, M.~Gavrilov, and J.~Bechhoefer.
\newblock High-precision test of {Landauer's} principle.
\newblock {\em Phys. Rev. Lett.}, 113:190601, 2014.

\bibitem{Boyd17a}
A.~B. Boyd, D.~Mandal, and J.~P. Crutchfield.
\newblock Thermodynamics of modularity: Structural costs beyond the landauer
  bound.
\newblock {\em Phys. Rev. X}, 8(031036), August 2018.

\bibitem{Kolc17}
A.~Kolchinsky and D.~H. Wolpert.
\newblock Dependence of dissipation on the initial distribution over states.
\newblock {\em J. Stat. Mech. Th. Exp.}, 2017(8):083202, 2017.

\bibitem{Engl13}
J.~L. England.
\newblock Statistical physics of self-replication.
\newblock {\em J. Chem. Physics}, 139(12):121923, 2013.

\bibitem{Wims19a}
G.~Wimsatt, O.-P. Saira, A.~B. Boyd, M.~H. Matheny, S.~Han, M.~L. Roukes, and
  J.~P. Crutchfield.
\newblock Harnessing fluctuations in thermodynamic computing via time-reversal
  symmetries.
\newblock {\em arXiv:1906.11973}.

\bibitem{Wims20a}
G.~W. Wimsatt, A.~B. Boyd, P.~M. Riechers, and J.~P. Crutchfield.
\newblock Balancing error and dissipation when erasing information.
\newblock {\em arXiv:2002.XXXXX}.

\bibitem{Shiv02}
P.~Shivakumar, M.~Kistler, S.~W. Keckler, D.~Burger, and L.~Alvisi.
\newblock Modeling the effect of technology trends on the soft error rate of
  combinational logic.
\newblock In {\em Dependable Systems and Networks, 2002. DSN 2002. Proceedings.
  International Conference on}, pages 389--398. IEEE, 2002.

\bibitem{Fred82b}
E.~Fredkin and T.~Toffoli.
\newblock Conservative logic.
\newblock {\em Intl. J. Theo. Phys.}, 21(3/4):219--253, 1982.

\bibitem{Fred02}
E.~Fredkin and T.~Toffoli.
\newblock Conservative logic.
\newblock In {\em Collision-based computing}, pages 47--81. Springer, 2002.

\bibitem{Riec17}
P.~M. Riechers and J.~P. Crutchfield.
\newblock Fluctuations when driving between nonequilibrium steady states.
\newblock {\em J. Stat. Physics}, 168(4):873--918, 2017.

\bibitem{Ging16}
T.~R. Gingrich, J.~M. Horowitz, N.~Perunov, and J.~L. England.
\newblock Dissipation bounds all steady-state current fluctuations.
\newblock {\em Phys. Rev. Lett.}, 116:120601, Mar 2016.

\bibitem{Hopf74}
J.~J. Hopfield.
\newblock Kinetic proofreading: {A} new mechanism for reducing errors in
  biosynthetic processes requiring high specificity.
\newblock {\em Proc. Natl. Acad. Sci.}, 71(10):4135--4139, 1974.

\bibitem{Benn79}
Charles~H. Bennett.
\newblock Dissipation-error tradeoff in proofreading.
\newblock {\em Biosystems}, 11(2):85 -- 91, 1979.

\bibitem{Muru12}
A.~Murugan, D.~A. Huse, and S.~Leibler.
\newblock Speed, dissipation, and error in kinetic proofreading.
\newblock {\em Proc. Natl. Acad. Sci.}, 109(30):12034--12039, 2012.

\bibitem{Sart15}
P.~Sartori and S.~Pigolotti.
\newblock Thermodynamics of error correction.
\newblock {\em Phys. Rev. X}, 5:041039, Dec 2015.

\bibitem{Horo17}
J.~M. Horowitz, K.~Zhou, and J.~L. England.
\newblock Minimum energetic cost to maintain a target nonequilibrium state.
\newblock {\em Phys. Rev. E}, 95:042102, Apr 2017.

\bibitem{Ould17}
T.~E. Ouldridge, C.~C. Govern, and P.~R. ten Wolde.
\newblock Thermodynamics of computational copying in biochemical systems.
\newblock {\em Phys. Rev. X}, 7(2):021004, 2017.

\bibitem{Bril56a}
L.~Brillouin.
\newblock {\em Science and Information Theory}.
\newblock Academic Press, New York, 1956.

\bibitem{Stei77a}
K.-L. Stein.
\newblock Noise-induced error rate as limiting factor for energy per operation
  in digital {IC}s.
\newblock {\em IEEE J. Solid State Circuits}, SC-12(5):527--530, 1977.

\bibitem{Mull76}
R.~Muller, H.J. Pfleiderer, and K.U. Stein.
\newblock Energy per logic operation in integrated circuits: Definition and
  determination.
\newblock {\em IEEE J. Solid-State Circ.}, 11(5):657--661, 1976.

\bibitem{Neum56a}
J.~von Neumann.
\newblock Probabilistic logics and the synthesis of reliable organisms from
  unreliable components.
\newblock In C.~E. Shannon and J.~McCarthy, editors, {\em Automata Studies},
  number~34 in Annals of Mathematical Studies, pages 329--378. Princeton
  University Press, Princeton, New Jersey, 1956.

\bibitem{Beni82a}
Paul~A. Benioff.
\newblock Quantum mechanical hamiltonian models of discrete processes that
  erase their own histories: Application to turing machines.
\newblock {\em Intl. J. Theo. Phys.}, 21(3-4):177--201, 1982.

\bibitem{Zure84}
W.~H. Zurek.
\newblock Reversibility and stability of information processing systems.
\newblock {\em Physical Review Letters}, 53(4):391, 1984.

\bibitem{Pere85}
A.~Peres.
\newblock Reversible logic and quantum computers.
\newblock {\em Phys. Rev. A}, 32:3266--3276, Dec 1985.

\bibitem{Lan12}
G.~Lan, P.~Sartori, S.~Neumann, V.~Sourjik, and Y.~Tu.
\newblock The energy--speed--accuracy trade-off in sensory adaptation.
\newblock {\em Nature Physics}, 8(5):422, 2012.

\bibitem{Zulk14a}
P.~R. Zulkowski and M.~R. DeWeese.
\newblock Optimal finite-time erasure of a classical bit.
\newblock {\em Phys. Rev. E}, 89:052140, 2014.

\bibitem{Lahi16}
S.~Lahiri, J.~Sohl-Dickstein, and S.~Ganguli.
\newblock A universal tradeoff between power, precision and speed in physical
  communication.
\newblock {\em arXiv:1603.07758}.

\bibitem{Deff2013}
S.~Deffner and C.~Jarzynski.
\newblock Information processing and the second law of thermodynamics: An
  inclusive, {Hamiltonian} approach.
\newblock {\em Phys. Rev. X}, 3:041003, 2013.

\bibitem{Owen17}
J.~A. Owen, A.~Kolchinsky, and D.~H. Wolpert.
\newblock Number of hidden states needed to physically implement a given
  conditional distribution.
\newblock {\em New J. Physics}, 21:013022, 2018.

\end{thebibliography}
\end{document}